\documentclass[11pt,a4paper]{article}
\usepackage{jheppub}

\usepackage{physics}
\usepackage{amsthm}
\usepackage{tikz}
\usepackage{float}
\usepackage{subfigure}
\usepackage{circuitikz}
\usepackage{longtable}
\usepackage{appendix}
\usepackage{pdfpages}
\usepackage{multirow}

\theoremstyle{plain}

\theoremstyle{definition}

\theoremstyle{remark}

\title{Holographic multipartite entanglement dynamics in AdS$_3$-Vaidya}

\author[a]{Xin-Xiang Ju,}
\author[b]{Ya-Wen Sun,}
\author[b]{Yang Zhao}
\author[b,c]{and Hao-Ran Zhou}

\affiliation[a]{Institute for Advanced Study, Tsinghua University, Beijing 100084, China}
\affiliation[b]{School of Physical Sciences, University of Chinese Academy of Sciences, Zhongguancun east road 80, Beijing 100190, China}
\affiliation[c]{Institute of Theoretical Physics, Chinese Academy of Sciences, Beijing 100190, China}
\emailAdd{{juxinxiang21@mails.ucas.ac.cn}, {yawen.sun@ucas.ac.cn}, {zhaoyang20a@mails.ucas.ac.cn}, 
{zhouhaoran22@mails.ucas.ac.cn}}

\abstract{We study how multipartite entanglement is dynamically reorganized during holographic thermalization following a global quench in AdS\(_3\)/CFT\(_2\). We first use the \(n\)-partite information \((-1)^n I_n\) to probe collective multipartite entanglement in holographic configurations where the full \(n\)-region entanglement wedge is connected while all fewer-party ones are disconnected, thereby excluding fewer-party contributions. The spatial range of multipartite entanglement first expands and then contracts as the system approaches its late-time locally thermal state. Entanglement involving different numbers of parties develops on comparable early-time scales, while the entanglement that involves more parties relaxes more slowly, revealing a transient propagation from shorter to longer spatial distances. We further compute the Markov gap and the genuine tripartite multi-entropy as complementary probes of tripartite entanglement. The Markov gap can remain enhanced after local thermalization, whereas the genuine tripartite multi-entropy undergoes a nonmonotonic evolution and returns to its vacuum value for the adjacent tripartition considered in this work. These results show that a global quench redistributes the entanglement across spatial scales and reorganizes its multipartite structure.}

\begin{document}
\maketitle

\section{Introduction}

{Thermalization is one of the central dynamical processes in quantum many-body physics: an initially far-from-equilibrium state evolves under unitary microscopic dynamics toward a state whose local observables are 
well 
described by a small number of thermodynamic parameters. This raises the basic question of how irreversible
macroscopic behavior emerges from reversible quantum evolution. In chaotic quantum systems, this question is
closely tied to the spreading and reorganization of quantum information: under unitary evolution, the
full state remains pure, while a small subsystem can nevertheless approach a thermal
reduced state as it becomes entangled with its complement \cite{Popescu:2006rhr,Goldstein:2005aib}.
Entanglement therefore provides a natural diagnostic of how thermal behavior is built up dynamically.
Holography~\cite{Maldacena:1997re} offers a particularly useful framework for studying this problem: it
provides access to strongly coupled quantum matter far from equilibrium, while entanglement quantities 
admit a
sharp geometric description in terms of extremal surfaces in the dual gravitational
spacetime~\cite{Ryu:2006bv,Ryu:2006ef,Hubeny:2007xt}.
}

{In holographic studies, a standard way to probe thermalization dynamically is to prepare the system in a
nonequilibrium state through a quantum quench and follow its subsequent real-time evolution. In this work, 
we
focus on a global quench in a holographic CFT,
whose bulk dual is modeled by an AdS-Vaidya geometry describing gravitational collapse
and black hole formation. This setup provides a controlled arena in which the boundary
state evolves from a far-from-equilibrium configuration toward local thermal
equilibrium, while the corresponding bulk extremal surfaces probe the time-dependent
geometry. Entanglement observables are particularly useful in this setting because
they are defined throughout the evolution and possess direct geometric duals.}

{Previous works have extensively investigated the real-time evolution of holographic
entanglement entropy, mutual information, and related entropy combinations in
AdS-Vaidya backgrounds~\cite{Abajo-Arrastia:2010ajo,
Balasubramanian:2011ur,Balasubramanian:2010ce,
Balasubramanian:2011at,Allais:2011ys,
Hartman:2013qma,Liu:2013iza,Liu:2013qca} and some other related backgrounds \cite{Erdmenger:2017gdk,Maulik:2025blg}. These studies established a characteristic
picture of scale-dependent thermalization: short-distance probes equilibrate earlier
than long-distance ones. The entanglement entropy of sufficiently large regions
exhibits an extended regime of approximately linear growth before saturation, and
large regions develop a thermal volume-law contribution associated with the formation
of a black hole horizon. Competing extremal surfaces can also produce nonmonotonic
behavior and sharp transitions in mutual and multipartite information. The resulting
entanglement-tsunami picture provides a useful description of how quantum information
spreads after a global quench.}

{Nevertheless, entanglement entropy is intrinsically bipartite: it measures the entropy
of a region relative to its complement, but does not resolve how correlations are
organized among the constituent subregions. In particular, the growth of entropy
during thermalization does not by itself determine whether the generated entropy is
stored as internal multipartite entanglement among spatial subregions or instead as
correlations with the exterior degrees of freedom that purify them. This motivates the
study of multipartite entanglement structures during holographic thermalization. By
decomposing the boundary into finer spatial subregions, one can track how multipartite
correlations spread, reorganize, and are eventually suppressed by thermal screening.
Such multipartite quantities provide a more refined probe of thermalization than entanglement
entropy alone, allowing one to distinguish the growth of entropy from changes in its
internal multipartite organization. In this sense, multipartite entanglement probes
not only how much entropy is generated during holographic thermalization, but also
where that entropy is encoded.}

{The dynamics of multipartite entanglement has only recently begun to be studied more
directly \cite{Balasubramanian:2024ysu}.  \cite{Balasubramanian:2025jhq} employed the entanglement-membrane
description to analyze several multipartite signals, including the residual information, tripartite
information, and
higher-party quantities, in the late-time and large-region regime of chaotic quantum many-body systems.
This approach revealed transient, nonmonotonic, and in some cases
discontinuous multipartite signal behavior, but it does not resolve the complete finite-time evolution in a
specific AdS-Vaidya geometry. More recently, \cite{Fujiki:2026qdt} studied the
genuine tripartite multi-entropy following a heavy local quench in a
holographic CFT$_2$. That analysis focused on
adjacent intervals and exploited the special description of the local-quench geometry
in terms of conical-defect or BTZ quotient geometries, where the dynamics is governed
by transitions between winding sectors. These developments, however, leave the full dynamical organization 
of multipartite entanglement throughout a global holographic quench largely unexplored.}

{In this work, we address this question by tracking the behavior of the multipartite entanglement among 
finer
spatial subregions in the global quench in AdS$_3$/CFT$_2$ dual to an AdS-Vaidya spacetime, from the initial
nonequilibrium state to equilibration. More specifically, we investigate the evolution of three types of 
multipartite entanglement quantities that detect distinct multipartite entanglement structures.
First, we consider the $n$-partite information for different numbers of disjoint boundary intervals 
\cite{Hayden:2011ag,Alishahiha:2014jxa}. Since the $n$-partite information is not a faithful measure of 
genuine 
multipartite entanglement in general \cite{Ju:2023tvo}, we study a controlled class of configurations in 
which 
its physical interpretation is sharpened: the holographic exclusive global multipartite entanglement 
configurations (HEGMECs) \cite{Ju:2025tgg,Ju:2024kuc,Ju:2024hba}. In an $n$-party HEGMEC, the full
entanglement wedge is connected whereas the wedges associated with all proper subsets are disconnected, 
thereby 
eliminating fewer-party contributions and allowing $(-1)^n I_n$ to serve as a clean measure of exclusive 
$n$-partite entanglement. Second, we study the Markov gap \cite{Dutta:2019gen,Hayden:2021gno} as a
complementary probe of
tripartite entanglement structure. A
nonzero Markov gap measures the obstruction to reducing a state,
after resolving the local Hilbert spaces into finer factors, to a
triangle state or a sum-of-triangle-states (SOTS) state \cite{Zou:2020bly,Akers:2019gcv}. Third, we compute 
the
genuine tripartite multi-entropy \cite{Gadde:2022cqi,Iizuka:2025ioc,Iizuka:2025caq}, obtained by subtracting
the bipartite
contributions from the tripartite multi-entropy. Together these observables probe the spatial range of 
multipartite
entanglement, as well as how their internal organization changes
during holographic thermalization and how the locally thermal entropy
of finite regions is purified by the remaining degrees of freedom
of the global pure state.}

We find that multipartite correlations are dynamically reorganized
across both spatial scales and structural classes. In the HEGMEC
regime, the collective \(n\)-partite signals are transiently enhanced
and extend over larger separations before being reduced by late-time
thermal screening. This provides a direct holographic picture of the
propagation of entanglement from shorter to longer spatial scales.
The Markov gap exhibits a different behavior: for sufficiently close
regions, its late-time value can remain above the vacuum one,
indicating that local thermalization does not necessarily make the
tripartite purification structure more reducible to pairwise connections. In contrast,
for the adjacent tripartition studied in the multi-entropy analysis,
the genuine tripartite multi-entropy undergoes a nonmonotonic transient evolution, but eventually returns to
its vacuum value. 
Our results on the spatial propagation of collective correlations and on the reorganization of irreducible 
tripartite structure reveal aspects of the quench dynamics beyond those captured by the previously studied 
growth of thermal entropy.

This paper is organized as follows. Section \ref{sec:vaidya_entropy} reviews the holographic entanglement 
entropy in the AdS$_3$-Vaidya geometry after a global quench. In Section \ref{sec3}, we introduce the upper 
bound of $(-1)^nI_n$ in holography, and show that $(-1)^nI_n$ becomes a good multipartite entanglement 
measure 
in the upper 
bound configurations, as well as the generalized HEGMEC configurations. Then we study $(-1)^nI_n$ in the 
HEGMEC 
configuration in the Vaidya geometry. 
% In Section \ref{sec4},
% we compute the Markov gap in AdS$_3$-Vaidya, while Section \ref{sec5} studies the genuine tripartite 
%multi-
% entropy in the Vaidya geometry. 
Subsequently, in Section \ref{sec4} and Section \ref{sec5}, the Markov gap
and the genuine tripartite multi-entropy in AdS$_3$-Vaidya are computed and analyzed. All these results are
used to analyze the time evolution of multipartite entanglement structures during thermalization. Section
\ref{sec6} concludes the paper.

\section{Holographic entanglement entropy in the AdS$_3$-Vaidya quench}
\label{sec:vaidya_entropy}

In this section, we review the holographic description of entanglement
entropy after a global quench in AdS$_3$ \cite{Abajo-Arrastia:2010ajo,Balasubramanian:2011ur}.
The quench is modeled by an AdS$_3$-Vaidya spacetime describing the collapse
of a homogeneous null shell and the subsequent formation of a planar BTZ black hole. 
% The purpose is to establish the basic method for calculating entanglement entropy using the HRT formula in
%the Vaidya geometry, which will be utilized  
The results obtained through the HRT formula in this section will be
utilized in the calculation of multipartite entanglement quantities in the following sections.

\subsection{The HRT prescription and the AdS$_3$-Vaidya geometry}

% Consider a quantum system with a Hilbert space
% $\mathcal H=\mathcal H_A\otimes\mathcal H_{\bar A}$ and density matrix
% $\rho$. The reduced density matrix and the von Neumann entropy associated
% with a subsystem $A$ are
% \begin{equation}
%     \rho_A=\operatorname{Tr}_{\bar A}\rho,
%     \qquad
%     S(A)=-\operatorname{Tr}\rho_A\log\rho_A .
% \end{equation}
% The entropy $S(A)$ measures the entanglement between $A$ and its complement
% when the full state is pure. More generally, for a mixed state it also
% contains contributions from classical correlations and thermal mixing \cite{Rangamani:2016dms}.\zy{this
%paragraph not necessary} \sun{yes, delete it}

For a static holographic state, the Ryu-Takayanagi prescription associates the entanglement entropy
$S(A)$ of boundary subregion $A$ with a codimension-two minimal surface $\gamma_A^{\rm RT}$ lying on a
static bulk time slice~\cite{Ryu:2006bv,Ryu:2006ef}. 
In a time-dependent geometry, the entanglement entropy is given by the covariant Hubeny-Rangamani-Takayanagi
prescription, which replaces the minimal surface by a spacelike extremal surface $\gamma_A$ in the full
Lorentzian spacetime~\cite{Hubeny:2007xt}
\begin{equation}
    S(A)=\frac{\operatorname{Area}(\gamma_A)}{4G_N},\quad \text{with}\quad \partial\gamma_A=\partial A,
    \quad
    \gamma_A\sim A.
    \label{eq:HRT_general}
\end{equation} 
Operationally, one first finds
all spacelike extremal surfaces satisfying these boundary and homology
conditions and then selects the candidate with the smallest area. In a
three-dimensional bulk, $\gamma_A$ is a spacelike geodesic and its area is
simply its proper length.

We consider a homogeneous global quench whose gravitational dual is the
collapse of a thin shell of null matter \cite{Abajo-Arrastia:2010ajo}. In ingoing
Eddington-Finkelstein coordinates, the AdS$_3$-Vaidya metric is
\begin{equation}
    ds^2
    =
    \frac{1}{z^2}
    \left[
       -f(v,z)\,dv^2
       -2\,dv\,dz
       +dx^2
    \right],
    \qquad
    f(v,z)=1-m(v)z^2 ,
    \label{eq:vaidya_ads_g}
\end{equation}
where the AdS radius has been set to unity. The coordinate $v$ is an ingoing
null coordinate. Near the asymptotic boundary, $t_b=\lim_{z\rightarrow 0}v$ is the physical time of the
boundary CFT, which we denote by $t_b$ in order to distinguish it from the static time coordinates 
introduced
separately in the AdS and BTZ regions.

The function $m(v)$ specifies the energy injected by the quench. We take the
thin-shell limit
\begin{equation}
    m(v)=M\,\Theta(v),
    \label{eq:thin_shell_mass}
\end{equation}
where $\Theta(v)$ is the Heaviside function. The null hypersurface $v=0$
represents an infinitesimally thin shell sent into the bulk from the
boundary at $t_b=0$. Since the shell is homogeneous along $x$, the quench
injects energy uniformly into the boundary system.

The geometry before the shell ($v<0$) is a Poincar\'e AdS$_3$ spacetime that describes the initial vacuum
state. The region
behind the shell ($v>0$)
is a planar BTZ black hole that describes the local equilibrium geometry
reached after the energy injection. It is convenient to introduce the inverse radial coordinate $r={1}/{z}$.
The full Vaidya metric then becomes
\begin{equation}
    ds^2
    =
    -\left[r^2-m(v)\right]dv^2
    +2\,dv\,dr
    +r^2dx^2 .
    \label{eq:vaidya_r}
\end{equation} Defining $r_H=\sqrt{M}$,
the final black hole temperature is
\begin{equation}
    T=\frac{r_H}{2\pi},
    \qquad
    \beta=\frac{2\pi}{r_H}.
\end{equation}
Thus $r_H$ determines the thermal length scale of the final state.

A key advantage of this thin-shell geometry is that the spacetime is static on
either side of the shell, hence different portions of a crossing
HRT geodesic can be solved analytically in the pure-AdS and BTZ regions and
then joined by the appropriate matching conditions at $v=0$. 

In the BTZ region, the static Schwarzschild time
$\tau_{\mathrm B}$ is introduced through
\begin{equation}
    v
    =
    \tau_{\mathrm B}
    +
    \frac{1}{2r_H}
    \log\left|
       \frac{r-r_H}{r+r_H}
    \right|.
    \label{eq:btz_EF_static}
\end{equation}
The metric takes the standard static form
\begin{equation}
    ds^2
    =
    -\left(r^2-r_H^2\right)d\tau_{\mathrm B}^2
    +
    \frac{dr^2}{r^2-r_H^2}
    +
    r^2dx^2 .
    \label{eq:BTZ_static}
\end{equation}
Similarly, in the pure-AdS region the Poincar\'e time
$\tau_{\mathrm A}$ is defined by
\begin{equation}
    v=\tau_{\mathrm A}-\frac{1}{r}.
    \label{eq:ads_EF_static}
\end{equation}
Both static time coordinates approach the boundary time:
\begin{equation}
    \tau_{\mathrm A},\tau_{\mathrm B}
    \longrightarrow t_b
    \qquad
    \text{as }r\rightarrow\infty.
\end{equation}
These distinctions are important because a shell-crossing HRT geodesic is
not a constant-$v$ curve and, in general, is not a constant-$\tau_{\mathrm
B}$ curve either. 
% \zy{Please confirm: from key advantage to here, put in appendix or keep it here? $\tau$
% seems rarely used.}

\subsection{HRT {surfaces} in the AdS$_3$-Vaidya geometry}\label{sec2.2}

% \zy{I think that in 2.2, maybe we should keep (2.40), and describe the definition of other elements (for
%example $r_c,r_\star$) used in the calculation of section 3,4,5 in one or two sentences. The specific
%calculation (that are not already known from previous Vaidya works) can be put into appendix.}
% \sun{yes! let's do this: we only make a brief state that there are three regimes and then give the result
%for the three regimes(2.39 now? the Lren(l, tb) formula right?) and keep the basic background introduction 
%of
%all the parameters in the formula, details could be found in references? no need for appendix maybe...}

First, for the HRT surfaces of a single interval, consider 
\begin{equation}
    A=
    \left[-\frac{\ell}{2},\frac{\ell}{2}\right]
\end{equation}
on the boundary time slice $t_b$. The geometry explored by its HRT surface depends on both $\ell$ and $t_b$.
There are three qualitatively different regimes:
\begin{enumerate}
    \item For $t_b\leq0$, the HRT geodesic lies entirely in the initial
    pure-AdS region.
    \item For $0<t_b<\ell/2$, the HRT surface intersects the shell. It consists of three segments: the 
    central
    segment resides in pure AdS, while two outer segments are in the BTZ region.
    \item For $t_b\geq\ell/2$, the HRT geodesic is located entirely in the final BTZ
    region. Its length equals the thermal equilibrium result.
\end{enumerate}
The second regime is the genuinely dynamical regime, where the HRT surface probes both sides of the 
collapsing
shell and {therefore captures the transition from the initial state to the final equilibrium. } 

Since these boundary-anchored geodesics have universal UV divergence, we first introduce the cutoff
$r_{max}=\frac{1}{\epsilon}$ on endpoints of $A$. The  geodesic length is then renormalized as:
\begin{equation}
    L^{\mathrm{ren}}(\ell,t_b)
    =
    \lim_{r_{\max}\rightarrow\infty}
    \left[
       \mathcal L(\ell,t_b;r_{\max})
       -
       2\log(2r_{\max})
    \right].
    \label{eq:Lren_definition}
\end{equation}
The subtracted term is universal for the asymptotically AdS geometries considered here and cancels in all
balanced entropy combinations evaluated below. Hence we work directly with $L^{\mathrm{ren}}$.

{Based on \cite{Balasubramanian:2011ur}, the renormalized geodesic lengths for the three regimes are:
\begin{equation}
    L^{\mathrm{ren}}(\ell,t_b)
    =
    \begin{cases}
    2\log\left(\dfrac{\ell}{2}\right),
    & t_b\leq0,
    \\[4mm]
    2\log\left[
       \dfrac{\sinh(r_Ht_b)}
       {r_Hs(\ell,t_b)}
    \right],
    & 0<t_b<\dfrac{\ell}{2},
    \\[5mm]
    2\log\left[
       \dfrac{
          \sinh\left(\frac{r_H\ell}{2}\right)
       }{r_H}
    \right],
    & t_b\geq\dfrac{\ell}{2}.
    \end{cases}
    \label{eq:vaidya_length}
\end{equation}
For the second regime, $s(\ell,t_b)$ is defined as the dimensionless ratio $\frac{r_\star}{r_c}$. Here
$r_\star(\ell,t_b)$ is the turning point of the geodesic that characterizes the deepest bulk location 
reached
by the boundary-anchored geodesic. $r_c(t_b)$ is the shell-crossing radius, where the geodesic passes 
through the
thin shell and enters the pure-AdS region. The detailed derivation of \eqref{eq:vaidya_length} is shown in
Appendix \ref{appA}, where the dependences of $r_\star$ and $r_c$ on $(\ell,t_b)$ are shown in equations
\eqref{eq:s_chi_rho}-\eqref{eq:s_implicit}.}

{From \eqref{eq:vaidya_length}, the full entropy of $A$ can be directly obtained,
and we work with its UV-finite sector 
\begin{equation}
    S_{\mathrm{ren}}(\ell,t_b)
    =
    \frac{c_{\mathrm{CFT}}}{6}
    L^{\mathrm{ren}}(\ell,t_b)
\end{equation}
in the following calculations. \eqref{eq:vaidya_length} also makes the scale-dependent saturation time 
manifest.
An interval of size $\ell$ reaches its thermal entropy at 
\begin{equation} 
    t_{\mathrm{sat}}(\ell)=\frac{\ell}{2}. 
\label{eq:saturation_time} 
\end{equation} 
This admits a direct bulk interpretation: small intervals are
probed by shallow HRT geodesics which enter the post-quench region quickly, while larger intervals are 
probed
by geodesics extending deeper into the bulk, whose intersection with the shell lasts for a longer time. 
Thus,
short distance correlations thermalize earlier than long distance correlations.}

Once the single-interval geodesic length is known, a candidate HRT surface for a union of boundary intervals
can be specified by an admissible non-crossing pairing of the interval endpoints, with each connected pair
representing a boundary-anchored geodesic between these endpoints. The entropy is then determined by
minimizing the total geodesic length over all pairings compatible with the homology constraint. 
Specifically,
for a boundary region $A$ composed of $N$ disjoint intervals:
\begin{equation}
    A
    =
    \bigcup_{a=1}^{N}
    [x_{2a-1},x_{2a}],
    \
    x_1<x_2<\cdots<x_{2N},
\end{equation}
its entropy is given by\begin{equation}
    S_{\mathrm{ren}}(A,t_b)
    =
    \frac{c_{\mathrm{CFT}}}{6}
    \min_{\mathcal P\in\mathfrak P_A}
    \left[
       \sum_{(i,j)\in\mathcal P}
       L^{\mathrm{ren}}(x_j-x_i,t_b)
    \right],
    \label{eq:multi_interval_entropy}
\end{equation}
where $\mathfrak P_A$ denotes the set of admissible non-crossing pairings homologous to $A$. Therefore, the
time-dependent function $L^{\mathrm{ren}}(\ell,t_b)$ in
 \eqref{eq:vaidya_length}, together with the global minimization over
allowed endpoint pairings in  \eqref{eq:multi_interval_entropy} contains
all the geometric information required to compute $I_n$
during the AdS$_3$-Vaidya quench. During this quench, changes in the dominant pairing reveal the phase 
transitions of entanglement wedge connectivity.

\section{The HEGMEC and multipartite information in AdS$_3$-Vaidya}\label{sec3}

Having established the HRT prescription for single and multiple intervals in
the AdS$_3$-Vaidya geometry, we now analyze the evolution of multipartite
correlations during holographic thermalization. In this section, we focus on
the $n$-partite information evaluated in holographic exclusive global
multipartite entanglement configurations (HEGMECs) \cite{Ju:2025tgg,Ju:2024kuc,Ju:2024hba}. We first review
mutual
information and the $n$-partite information $I_n$ in Vaidya geometry. For a generic multipartite state,
however, $I_n$ is not a faithful
measure of genuine $n$-partite entanglement \cite{Ju:2023tvo,Ju:2024hba}: it can contain contributions
associated with fewer-party correlations, and for $n\geq3$ it does not possess
the general positivity and monotonicity properties expected of an entanglement measure. We therefore
specialize to HEGMECs, in which the  entanglement wedge of the full $n$-party system is connected while the
entanglement wedges of all proper subsets are disconnected. In this
configuration $(-1)^n I_n$ provides a clean diagnostic of the
collective irreducible entanglement that involves all $n$ regions. We use this quantity to
study how the spatial range and strength of multipartite correlations evolve
throughout the Vaidya quench.

\subsection{Mutual information and $n$-partite information}

For two subsystems $A$ and $B$, the mutual information is defined as
\begin{equation}
  I(A,B)=S(A)+S(B)-S(A \cup B),
\end{equation}
and captures the entanglement between $A$ and $B$. It can then be generalized
to an arbitrary number of subsystems to probe
correlations shared by these subsystems. For $n$ subsystems
$A_1,\ldots,A_n$, the $n$-partite information is defined from the
entropies of their unions as 
\begin{equation}
  I(A_1,\ldots,A_n)
  =
  \sum_{i=1}^{n} S(A_i)
  -
  \sum_{\substack{i<j \\ i,j=1,\ldots,n}}S(A_i \cup A_j)
  +\cdots
  +(-1)^{n-1}S(A_1 \cup \cdots \cup A_n).
\end{equation}
For example, when $n=3$ we have  \cite{Hayden:2011ag}
\begin{equation}
  I_{3}(A,B,C)=S(A)+S(B)+S(C)-S(A\cup B)-S(A\cup C)
               -S(B\cup C)+S(A\cup B\cup C).
\end{equation}

%\zy{This paragraph is not necessary?}Consider $n$ intervals of length $l$, with all adjacent separations
%given by
%$d$. For $n=2,3,4,5,6$, $I_n$ is given by the results for $S_n$ obtained above. For $n=3$, the explicit
%result of $I_3$ is
%\begin{equation} I_3(l,d,t)=3L(l,t)-2S_2(l,d,t)-S_2(l,l+2d,t) +S_3(l,d,d,t). \end{equation}

The real-time evolution of mutual and tripartite information in
AdS$_3$-Vaidya backgrounds has been studied in
\cite{Balasubramanian:2011at,Allais:2011ys}. For two disjoint intervals, the mutual information and 
tripartite
information typically exhibit a non-monotonic profile, increasing above its vacuum value before decreasing
toward the final thermal value. The analysis was subsequently extended to higher $n$-partite information,
including $I_4$ and $I_5$, for symmetric collections of parallel strips in AdS-Vaidya geometries
\cite{Alishahiha:2014jxa}. In the close-strip regime, these quantities display several distinct scaling 
stages
during thermalization.

These results demonstrate that $I_n$ is sensitive to the dynamical
reorganization of correlations. However, $I_n$ is in general not a good measure of genuine $n$-partite
entanglement. This is particularly transparent for three
parties. For any pure state on $ABC$,
\begin{equation}
    S(AB)=S(C),\qquad
    S(AC)=S(B),\qquad
    S(BC)=S(A),\qquad
    S(ABC)=0,
\end{equation}
and hence
\begin{equation}
    I_3(A:B:C)=0.
\end{equation} {$I_3$ is thus not a faithful measure: since $I_3$ vanishes for every pure tripartite state, 
it
cannot tell a fully separable state from genuinely tripartite-entangled states, or distinguish inequivalent
entanglement structures such as the GHZ and W
states~\cite{Greenberger:1990uox,Dur:2000zz,Walter:2016lgl}. More generally,
$I_n$ is a signed inclusion-exclusion combination of subsystem entropies. It therefore mixes correlations of
different numbers of parties. Their contributions may cancel in $I_n$ even when genuine $n$-partite
entanglement is present. For mixed states, the interpretation of $I_n$ is further complicated by possible
contributions from classical and fewer-party quantum correlations.}

\subsection{Upper-bound configurations and the HEGMEC in holography}

{The discussion above shows that $I_n$ is generally not a faithful measure of genuine multipartite
entanglement, since different types of correlations can contribute to the same entropy combination. 
Therefore,
additional constraints are needed to extract such genuine $n$-partite correlations from $I_n$. By requiring
$(-1)^n I_n$ to saturate its information-theoretic upper bound, we obtain a particularly important class of
configurations, where the multipartite correlations are maximal while contributions from fewer-partite and
classical correlations are removed \cite{Ju:2025tgg,Ju:2024kuc,Ju:2024hba}. Consequently, within this
restricted class of states, $(-1)^n I_n$ acquires a direct interpretation as the genuine $n$-partite quantum
contribution.}

{In particular, we fix $n-1$ boundary subregions $A_1,\cdots,A_{n-1}$, and adjust the $n$-th region $E$ so
that $(-1)^n I_n(A_1,\cdots:A_{n-1},E)$ reaches the upper bound in quantum information theory. Such a
configuration of $\{A_1,\cdots,A_{n-1},E\}$ is called an upper-bound configuration. In the tripartite case, 
we
fix $A,B$ and vary the position and size of $E$ to maximize $-I_3(A:B:E)$. Since
\begin{equation}
  -I_3(A:B:E)=I(A:B|E)-I(A:B),
\end{equation}
maximizing $-I_3$ is equivalent to maximizing the conditional mutual
information $I(A:B|E)$.}

In quantum information theory, the Araki-Lieb inequality gives \cite{araki1970entropy}
\begin{equation}
  I(A:B|E) \leq 2\min(S_A,S_B).
\end{equation}\label{QIbound}
Without loss of generality, we assume that $S(A)<S(B)$. 
Saturation requires that $A$ have no correlation with $B$ or with $E$ separately, while its correlation with
the union $BE$ reaches the maximal value \cite{Shirokov:2017zka},
\begin{equation}
  I(A:B)=0,\qquad I(A:E)=0,\qquad I(A:BE)=2S_A.
\end{equation} Moreover, all correlations contributing to $-I_3$ must be quantum in nature. Any classical
contribution would lead to a deviation from the upper bound.

{In \cite{Ju:2025tgg,Ju:2024kuc,Ju:2024hba}, we focus on the holographic realizations of the upper-bound
configurations. We 
established a bridge between the information-theoretic upper bound and the holographic  geometric structure 
of
entanglement, by characterizing these upper-bound saturation constraints in terms of entanglement-wedge
connectivity conditions of the boundary regions. Specifically, the upper bound of $(-1)^n 
I_n(A_1,\cdots:A_{n
},E)$ is only reached when $E$ is tuned in such a way that the full $n$-party entanglement wedge remains
connected, while the wedges associated with fewer-partite subsets are
disconnected. In this configuration, the fewer-party correlations are subtracted due to the disconnected
wedges, while the sole contribution of $n$-partite entanglement constitutes the maximal value of $(-1)^n 
I_n$.
The upper-bound configuration thus provides an ideal example in which
$(-1)^nI_n$ has a direct multipartite entanglement interpretation.}

{Motivated
by this observation, we next consider a broader holographic class of
configurations, namely the holographic exclusive global multipartite
entanglement configurations (HEGMECs) \cite{Ju:2025tgg}. The HEGMECs are defined as the configurations where
the entanglement wedge of
the full $n$-party system is connected, while all its proper subsets have disconnected entanglement wedges.
Although they do not necessarily saturate the information-theoretic upper bound, the exclusion of all fewer
party connected wedges allows $(-1)^nI_n$ to isolate the collective $n$-party correlation in a geometrically
transparent way{, and therefore to serve as a well-defined diagnostic of the entanglement shared by all $n$
subsystems. In the next subsection, we use HEGMECs to probe the evolution of multipartite entanglement
structures during the AdS$_3$-Vaidya quench, compare their behavior in the initial vacuum and the final
thermal state, and give a special analysis of the configurations in which $(-1)^n I_n$ reaches its 
information
theoretic upper bound.}}

\subsection{$(-1)^nI_n$ on HEGMEC in AdS$_3$-Vaidya}
\label{subsec:In_HEGMEC_Vaidya}

Having introduced HEGMECs and clarified the interpretation of $(-1)^n I_n$ within these configurations, we 
now apply this construction to the AdS$_3$-Vaidya quench. We consider a family of disjoint intervals and 
explicitly determine the HRT surfaces entering the $n$-partite information. Since $I_n$ sums over entropies 
of all union subsets, we first classify the minimal geodesic configurations. Each union allows multiple HRT 
topologies, with 
the dominant one varying during the quench. 

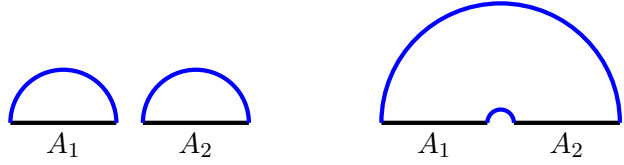
\begin{figure}[H]
\begin{center}
\begin{tikzpicture}[scale=0.7]
\draw[ultra thick,black] (0,0) -- (2,0);
\draw[ultra thick,black] (2.5,0) -- (4.5,0);
\draw[ultra thick,blue] (2,0) arc (0:180:1cm);
\draw[ultra thick,blue] (4.5,0) arc (0:180:1cm);
\draw[ultra thick,black] (7,0) -- (9,0);
\draw[ultra thick,black] (9.5,0) -- (11.5,0);
\draw[ultra thick,blue] (9.5,0) arc (0:180:0.25cm);
\draw[ultra thick,blue] (11.5,0) arc (0:180:2.25cm);
\draw[] (1,0) node[below] {$A_1$}; 
\draw[] (3.5,0) node[below] {$A_2$}; 
\draw[] (8,0) node[below] {$A_1$}; 
\draw[] (10.5,0) node[below] {$A_2$}; 
\end{tikzpicture}
\caption{Two candidate configurations to be considered when $n=2$.}
\label{fig:config2}
\end{center}
\end{figure}

Consider $n$ ordered intervals with length $l$ and separation $d_{(i)}$ between the $i$-th and the $i+1$-th
adjacent intervals. For $n=2$, there are two candidate
HRT configurations, as shown in Figure~\ref{fig:config2}. Their total lengths are 
\begin{align}
    \mathcal L_{2}^{\mathrm{dis}}(l,d,t)
    &=
    2L(l,t),
    \label{eq:L2_dis}
    \\
    \mathcal L_{2}^{\mathrm{con}}(l,d,t)
    &=
    L(d,t)+L(2l+d,t),
    \label{eq:L2_con}
\end{align}
where $t$ denotes the boundary time.
The entropy is therefore
\begin{equation}
    S_2(l,d,t)
    =
    \frac{c_{\mathrm{CFT}}}{6}
    \min
    \left\{
        \mathcal L_{2}^{\mathrm{dis}}(l,d,t),
        \mathcal L_{2}^{\mathrm{con}}(l,d,t)
    \right\}.
    \label{eq:S2_explicit}
\end{equation}
The two branches correspond respectively to the disconnected and connected
entanglement wedges. Their exchange of dominance determines the transition
between vanishing and non-vanishing mutual information.

For $n=3$, there
are five admissible non-crossing pairings. Their
total HRT geodesic lengths are:
\begin{align}
    \mathcal L_{3}^{(1)}
    &=
    3L(l,t),
    \label{eq:L3_1}
    \\
    \mathcal L_{3}^{(2)}
    &=
    L(l,t)
    +L(d_{(1)},t)
    +L(2l+d_{(1)},t),
    \label{eq:L3_2}
    \\
    \mathcal L_{3}^{(3)}
    &=
    L(l,t)
    +L(d_{(2)},t)
    +L(2l+d_{(2)},t),
    \label{eq:L3_3}
    \\
    \mathcal L_{3}^{(4)}
    &=
    L(d_{(1)},t)
    +L(d_{(2)},t)
    +L(3l+d_{(1)}+d_{(2)},t),
    \label{eq:L3_4}
    \\
    \mathcal L_{3}^{(5)}
    &=
    L(l,t)
    +L(l+d_{(1)}+d_{(2)},t)
    +L(3l+d_{(1)}+d_{(2)},t).
    \label{eq:L3_5}
\end{align}
Hence,
\begin{equation}
    S_3(l;d_{(1)},d_{(2)};t)
    =
    \frac{c_{\mathrm{CFT}}}{6}
    \min_{1\leq a\leq5}
    \left\{
        \mathcal L_{3}^{(a)}
    \right\}.
    \label{eq:S3_explicit}
\end{equation}
These five candidates represent the different possible connectivity patterns
of the three-interval entanglement wedge. Since their individual lengths
evolve differently in the Vaidya geometry, the dominant pattern can change
with time for fixed  boundary intervals.

More generally, the number of non-crossing pairings of $2n$ ordered endpoints is the Catalan number
\begin{equation}
    C_n
    =
    \frac{1}{n+1}
    \binom{2n}{n}.
\end{equation}
Thus, there are $14$, $42$, and $132$ candidate pairings for $n=4,5,$ and $6$, respectively 
\cite{Alishahiha:2014jxa,Mirabi:2016elb}. In the equal-separation  configuration used below,
\begin{equation}
    d_{(1)}=d_{(2)}=\cdots=d_{(n-1)}=d,
\end{equation}
different pairings can produce identical total-length expressions because their geodesic components have 
the same sets of endpoint separations. After identifying these degeneracies, the $132$ pairings for six 
intervals reduce to $47$ inequivalent candidate length expressions. In our numerical calculation, all 
admissible pairings are generated and compared before the minimum is selected. The explicit formulas for 
$n=4,5,$ and $6$ are lengthy and do not provide additional conceptual insight. In the following analysis, 
they are evaluated algorithmically using \eqref{eq:multi_interval_entropy}.

With the multi-interval entropies determined above, we can now construct the $n$-partite information and 
identify the parameter region in which the corresponding configuration is an HEGMEC. For fixed $(l,t)$, the 
connectivity of the entanglement wedge is determined by $d$: sufficiently close intervals have a connected 
collective wedge, while large enough separations result in a disconnected  configuration. We define 
$d_n(l,t)$ as the critical distance where $EW(A_1\cup\cdots\cup A_n)$ transitions between the connected 
($d<d_n(l,t)$) and disconnected ($d>d_n(l,t)$) phases. In this equal-interval configuration, a boundary 
union containing more intervals remains connected up to a larger separation distance. Therefore, the 
critical separation distances satisfy
\begin{equation}
    d_2(l,t)<d_3(l,t)<\cdots<d_n(l,t).
    \label{eq:dn_ordering}
\end{equation}

This ordering considerably simplifies the HEGMEC conditions. Among all proper subsets of an $n$-interval 
system, the $(n-1)$-interval subsets are the most likely to remain connected. It follows that the $n$-interval system forms an $n$-partite HEGMEC precisely in the window
\begin{equation}
    d_{n-1}(l,t)<d<d_n(l,t),
\label{eq:HEGMEC_window}
\end{equation}
which ensures that the entanglement wedge of the full $n$-interval union is connected, while the wedges of 
all proper subsets are disconnected. Consequently, within this window all $I_{m,m<n}$ vanish while only 
$I_n$ remains nonzero, and the irreducible entanglement involving all $n$ intervals is therefore isolated.

In the following, we determine the critical curves $d_n(l,t)$ for $n=2,\ldots,6$ throughout the AdS$_3$-Vaidya evolution. These curves reveal the existence and specify the spatial range of the corresponding 
HEGMECs. We then evaluate $(-1)^n I_n$ inside each interval $d_{n-1}<d<d_n$ and introduce the entanglement 
strength $E_n(t)$ to characterize the total magnitude of the $n$-partite signal over its allowed {HEGMEC-supporting} separation range.

\subsubsection{Behavior of $d_n(l,t)$}

\begin{figure}[H]
  \centering
  \subfigure[$l=1$\label{fig:dl1}]{
    \includegraphics[width=0.47\textwidth]{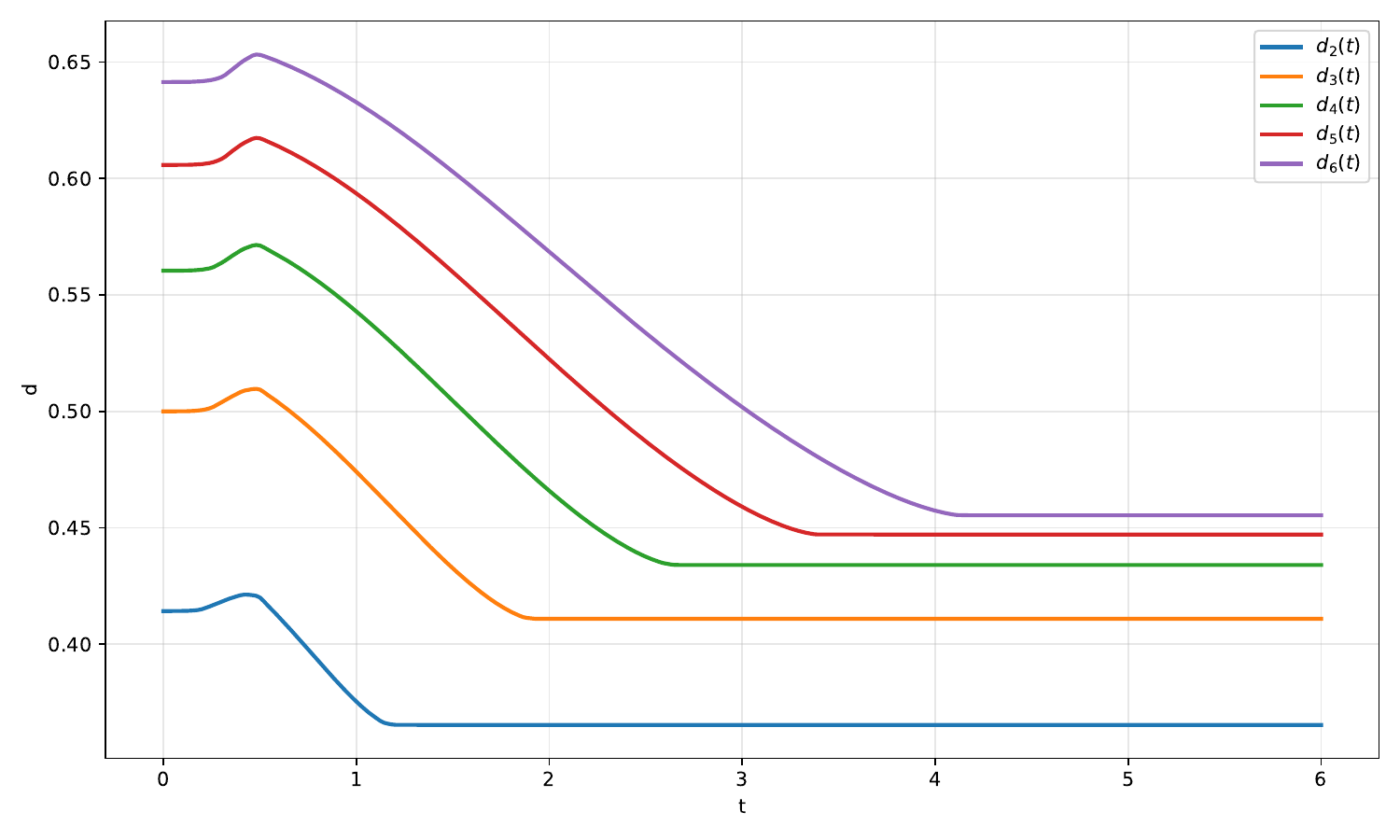}
  }\hfill
  \subfigure[$l=2$\label{fig:dl2}]{
    \includegraphics[width=0.47\textwidth]{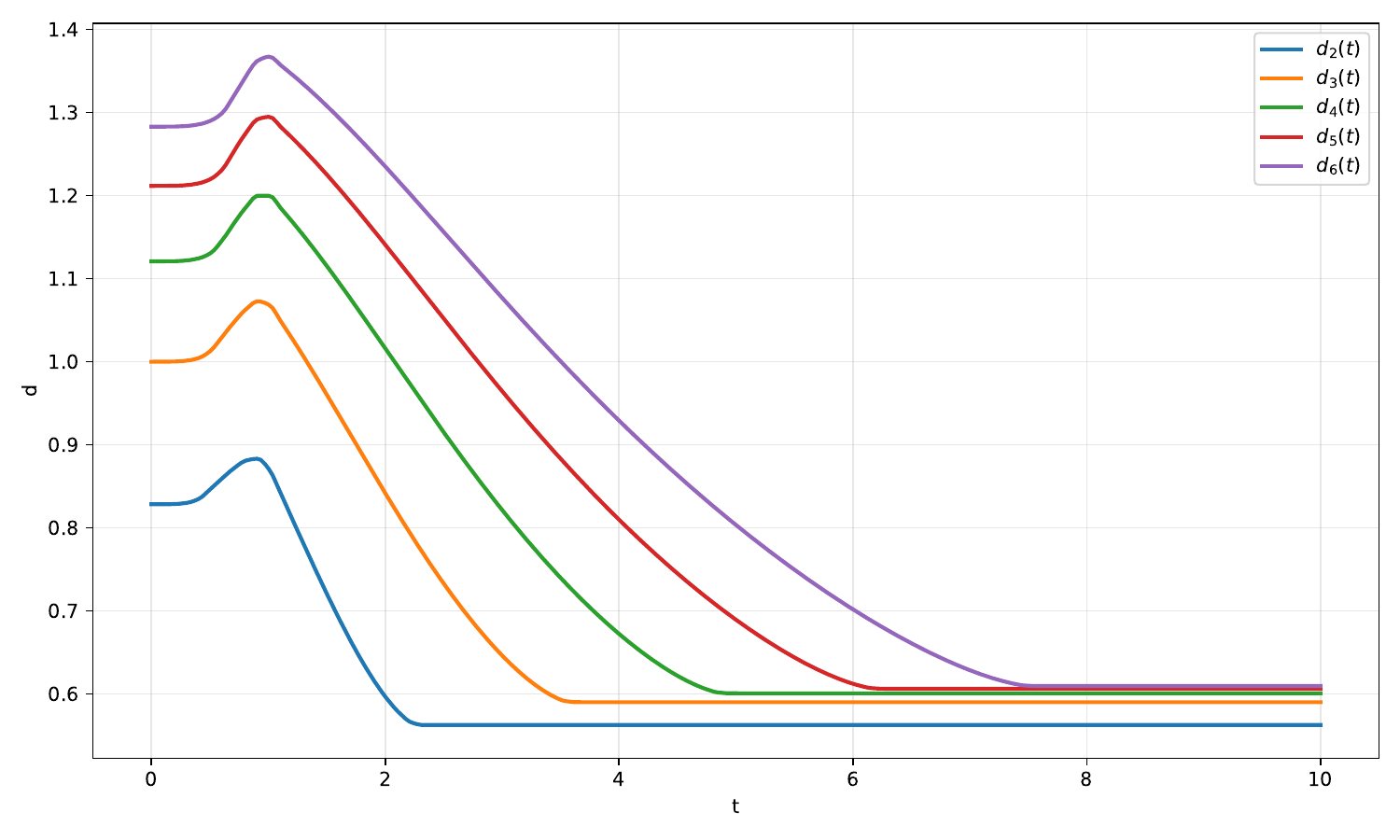}
  }\\[0.5em]
  \subfigure[$l=3$\label{fig:dl3}]{
    \includegraphics[width=0.47\textwidth]{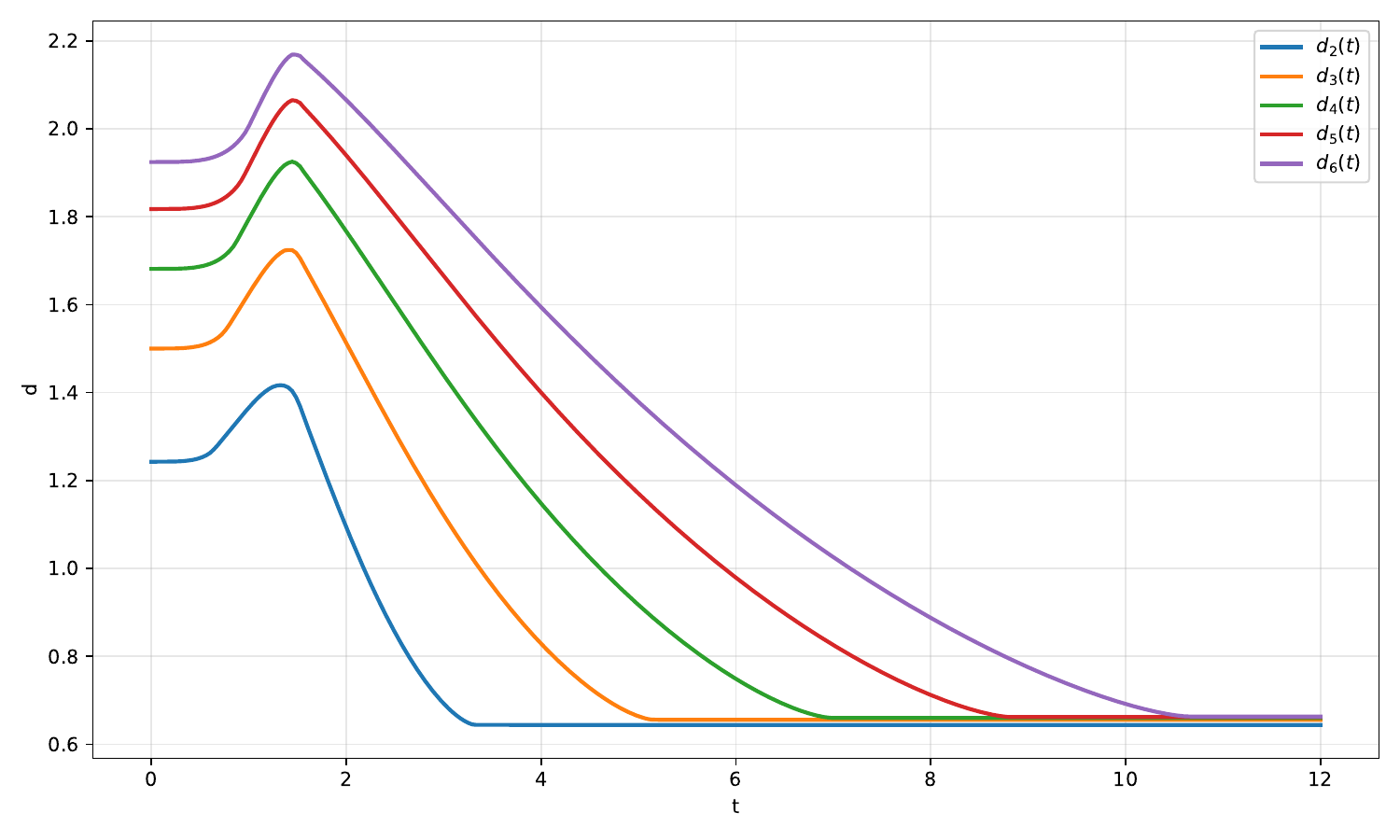}
  }\hfill
  \subfigure[$l=4$\label{fig:dl4}]{
    \includegraphics[width=0.47\textwidth]{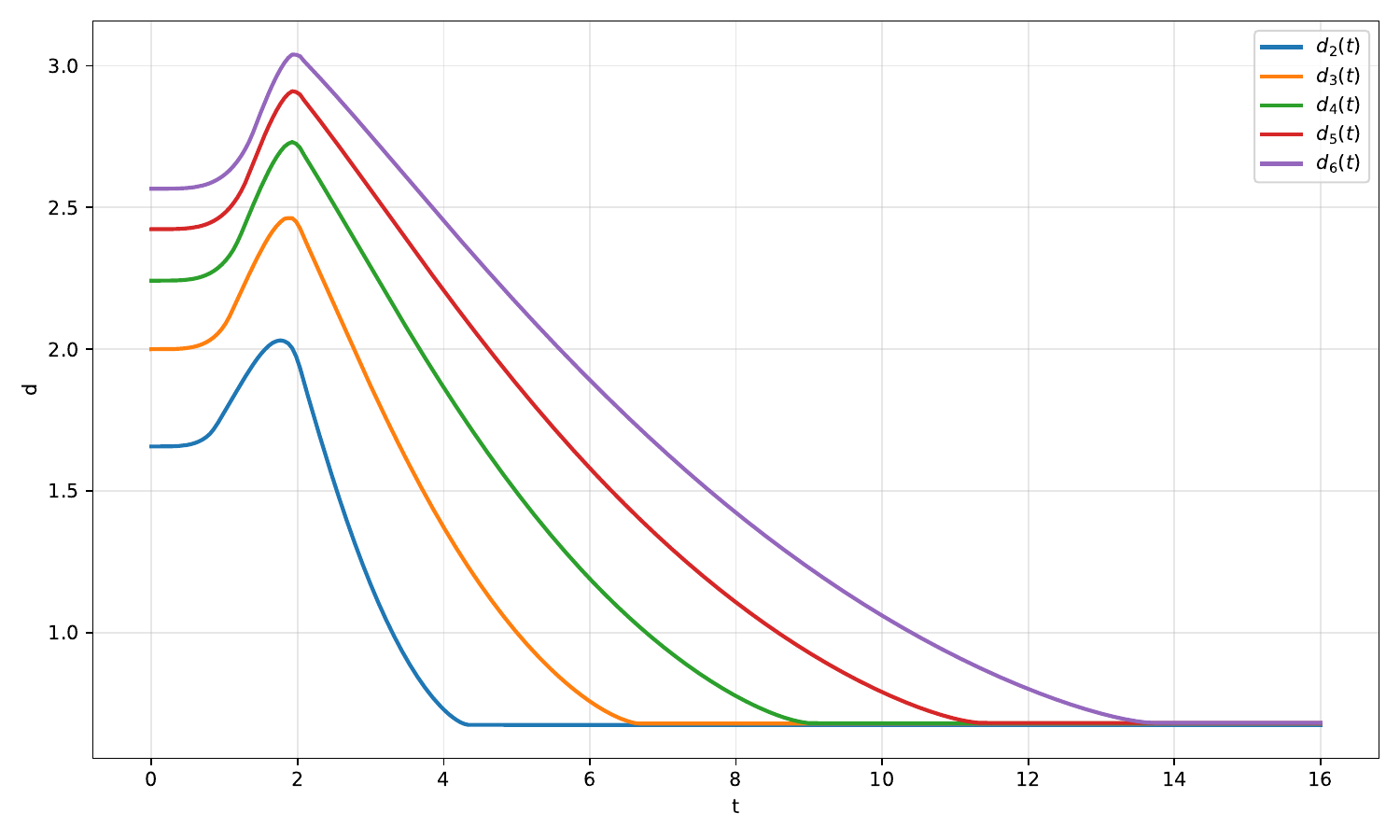}
  }\\[0.5em]
  \subfigure[$l=5$\label{fig:dl5}]{
    \includegraphics[width=0.47\textwidth]{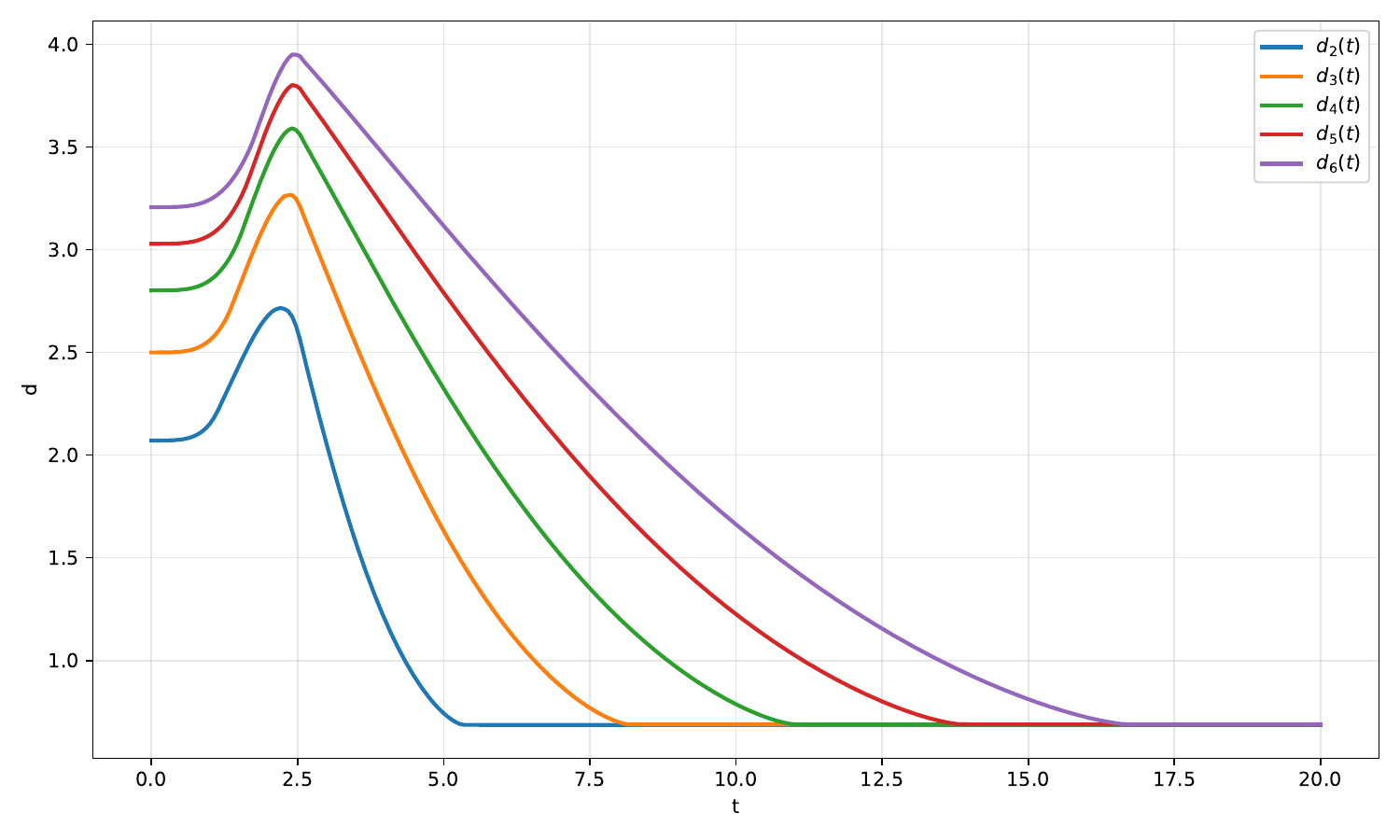}
  }\hfill
  \subfigure[$l=10$\label{fig:dl10}]{
    \includegraphics[width=0.47\textwidth]{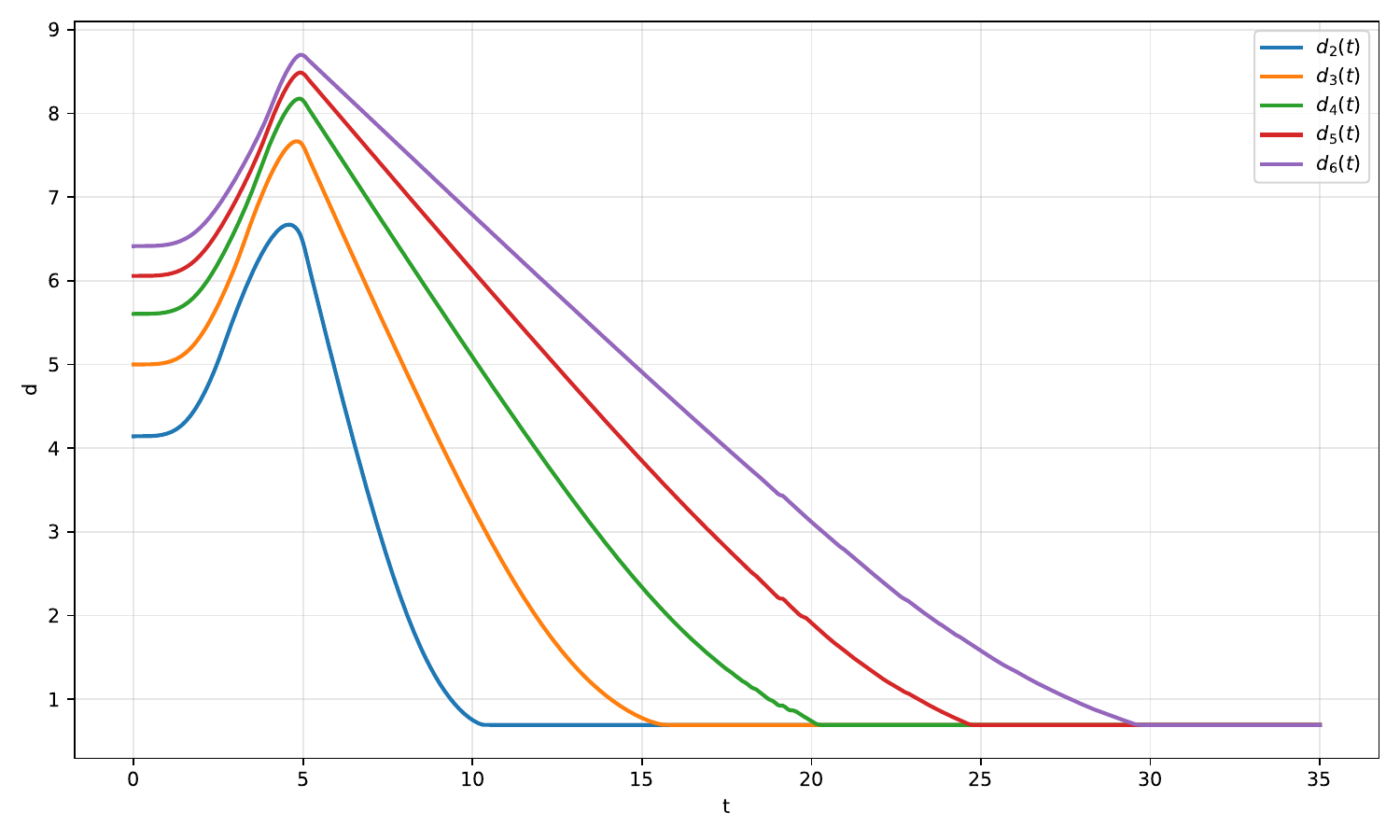}
  }
  \caption{Time evolution of $d_2(t),d_3(t),d_4(t),d_5(t),d_6(t)$ for $r_H=1$ and $l=1,2,3,4,5,10$. $n$
  subregions with equal lengths $l$ and separation $d$ form an $n$-party HEGMEC when $d_{n-1}(l,t)
  <d<d_n(l,t)$, and $I_n$ is nonzero in this window.}
  \label{fig:dls}
\end{figure}

Owing to the spatial ordering and equal separation of the intervals, the most strongly correlated subset of 
a given size is always a set of consecutive intervals. For example, the nearest-neighbor pair 
$(A_i,A_{i+1})$ has the smallest separation, so once its entanglement wedge becomes disconnected, the 
wedges of all more distant pairs become disconnected. The first critical curve is determined from
\begin{equation}
    I_2(l,d_2,t)=0,
\label{eq:d2_definition}
\end{equation}
where the equality denotes the transition between the connected and disconnected HRT configurations. The 
same simplification applies also to higher parties. We solve $I_n(l,d_n,t)=0$ numerically for 
$n=2,\ldots,6$. Setting $r_H=1$ and taking $l=1,2,3,4,5,$ and $10$, the resulting critical curves are shown 
in Figure~\ref{fig:dls}.

\begin{figure}[H]
  \centering
  \includegraphics[width=0.7\textwidth]{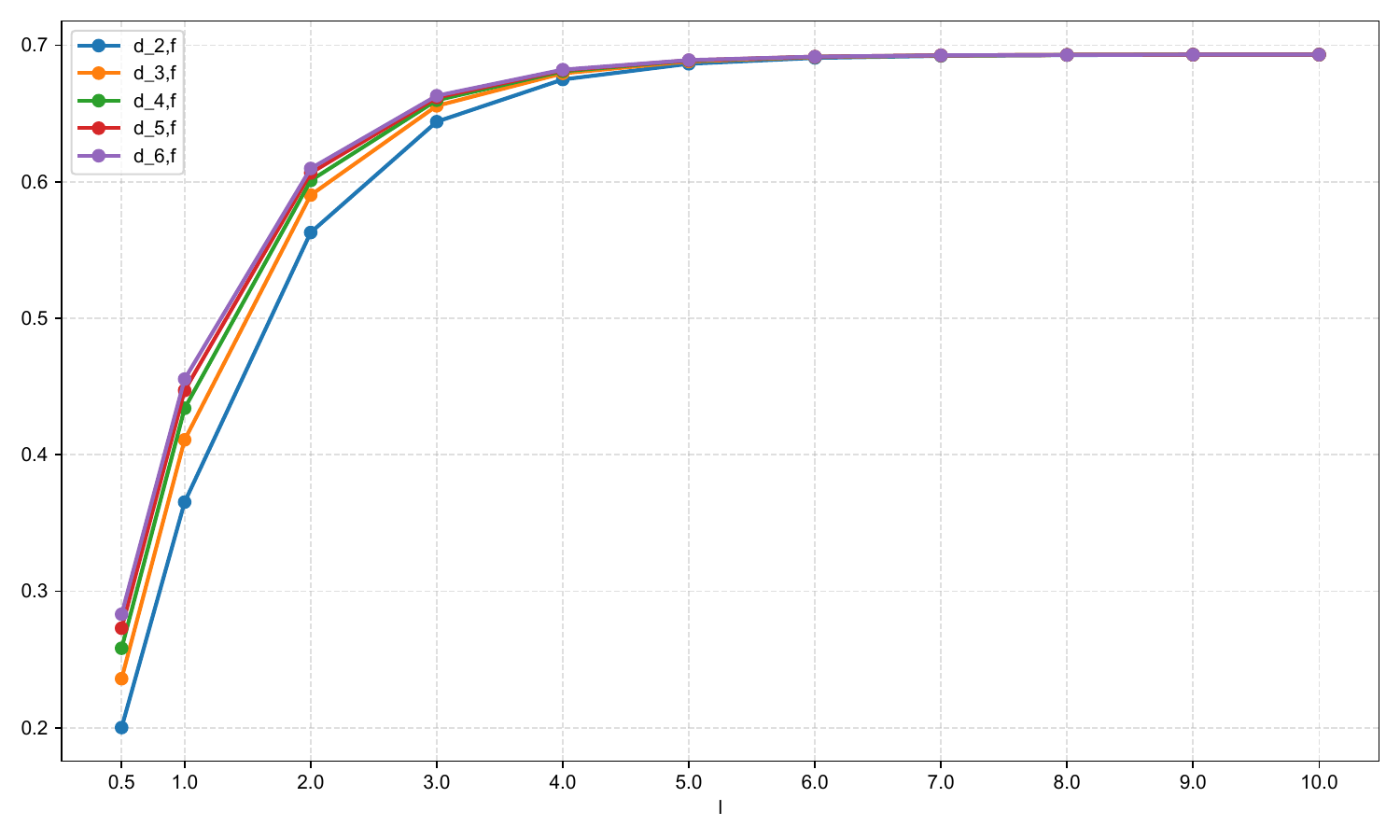}
  \caption{Dependence of the final values $d_{n,f}$ on $l$ for $r_H=1$.}
  \label{fig:df}
\end{figure}

As shown in Figure~\ref{fig:dls}, the critical separation distances exhibit a common
nonmonotonic evolution. We denote the maximum point that $d_n(l,t)$ reaches by
$(t_{n,\max},d_{n,\max})$, and the onset of the final plateau by
$(t_{n,f},d_{n,f})$. The dependence of the plateau value $d_{n,f}$ on the
interval length $l$ is displayed in Figure~\ref{fig:df}. In the following, we discuss three aspects of the
behavior of $d_n(l,t)$ as could be observed in the figures. 

{\noindent{\textbf{I. Nonmonotonic evolution and peak times of \(d_n\).}}

As shown in Figure~\ref{fig:dls}, for all values of \(n\) considered
and for different interval lengths \(l\), the critical separations
\(d_n(l,t)\) exhibit the same qualitative nonmonotonic evolution.
Starting from their vacuum values, all \(d_n\) first increase after the
quench, reach a single maximum \(d_{n,\max}\) at
\(t=t_{n,\max}\), and then decrease toward their late-time thermal
values \(d_{n,f}\). Since \(d_n(l,t)\) is the critical nearest-neighbor separation below which the 
entanglement 
wedge of the \(n\)-interval union remains connected, the initial increase of \(d_n\) indicates a transient 
enlargement of the separation distances where \(n\)-party global entanglement can exist. The subsequent 
decrease shows that this enlarged connectivity range is gradually reduced as the system approaches local 
thermal equilibrium. 

A notable feature is that the peak times \(t_{n,\max}\) for different
\(n\) are clustered within a narrow time window and are all slightly
earlier than the single-interval saturation time
\(
t_{\mathrm{sat}}(l)=\frac{l}{2}.
\) 
Therefore, by the time an interval of length \(l\) reaches its saturation of thermal
entropy, all the \(d_n\) curves have already entered their decreasing
regime. This near coincidence indicates that the transient growth of
the different \(n\)-region connectivity scales occurs approximately
synchronously and is governed primarily by the interval scale \(l\).

The late-time relaxation, however, is not synchronized. The fewer-party critical separations, such as 
\(d_2\), 
reach their final values earlier, whereas the curves with larger \(n\) remain above
their thermal plateaus for longer times. Namely, the buildup of the
connectivity ranges is approximately simultaneous, while their decay is hierarchical. The overshoot 
\(d_{n,\max}>d_{n,f}\) further shows that the enlarged connectivity
range is a transient nonequilibrium feature of the quench, whereas
the final plateau is determined by the thermal scale of the late-time
state.}

\noindent{ \textbf{II. Propagation of multipartite entanglement from shorter to longer-range.}}

The hierarchy of the critical curves provides a direct picture of how the spatial organization of 
multipartite entanglement changes during the quench. This can be seen by fixing the interval length and 
separation, which corresponds to drawing a horizontal line across the curves \(d_n(l,t)\). Each crossing of 
the curves then marks 
a change in the connectivity structure supported by that fixed spatial configuration.

As a first example, consider a boundary chain containing at least six intervals with \(l=5\) and \(d=2.2\), 
which we regard as a horizontal line between the coordinates $d=2.0$ and $d=2.5$ in Figure~\ref{fig:dls}
(e). At 
early times, the configuration lies in the window
\begin{equation}
    d_2(l,t)<d<d_3(l,t).
\end{equation}
All pairwise entanglement wedges are therefore disconnected, whereas
the wedge of three consecutive intervals remains connected. The
configuration is consequently a three-party HEGMEC, and \(-I_3\) measures the irreducible tripartite 
entanglement structure. Meanwhile, \(I_{n>3}\) are still
not good $n$-party measures in this window, since {nonzero tripartite entanglement contributions are mixed
with more-party correlations.}

During the subsequent growth regime, \(d_2(l,t)\) rises above the
fixed separation \(d\), and the nearest-neighbor mutual information
becomes temporarily nonzero. This shows that the correlation front
generated by the quench has reached the scale of the gap region between
adjacent intervals, thereby opening a pairwise connectivity channel
that was absent in the initial state. When \(d_2(l,t)\) later
decreases below \(d\), the pairwise wedges disconnect again, while the
three-interval wedge remains connected. The system therefore re-enters
the three-party HEGMEC window.

At still later times, the decreasing curves
\(d_3,d_4,d_5,\) and \(d_6\) successively cross the fixed separation.
The system consequently passes through the sequence
\begin{equation}
\begin{aligned}
    d_3<d<d_4
    &\quad &&\text{\(n=4\) HEGMEC},\\
    d_4<d<d_5
    &&&\text{\(n=5\) HEGMEC},\\
    d_5<d<d_6
    &&&\text{\(n=6\) HEGMEC}.
\end{aligned}
\end{equation}
Within each window, \((-1)^n I_n\) has a sharpened interpretation as
the exclusive collective correlation involving \(n\) consecutive
intervals, with all proper-subset entanglement wedges
disconnected. Eventually, when \(d>d_6(l,t)\), even the union of six
intervals becomes disconnected.

A complementary pattern appears for a larger fixed separation. Take
\(l=5\) as in Figure~\ref{fig:dls}(e) and fix the separating distance at
\begin{equation}
    d_4(l,0)<d<d_5(l,0).
\end{equation}
Initially, the configuration is a five-party HEGMEC: unions of four
or fewer consecutive intervals are disconnected, whereas the union
of five intervals remains connected. During the growth stage,
\(d_4(l,t)\) and subsequently \(d_3(l,t)\) can rise above the fixed
separation. The configuration can therefore pass temporarily through
four-party and three-party HEGMEC windows. By contrast,
\(d_2(l,t)\) never reaches this value of \(d\), so the pairwise
entanglement wedge remains disconnected and the nearest-neighbor
mutual information stays zero throughout the evolution.

This second example shows that the opening of fewer-party
connectivity channels is constrained. For a fixed separation \(d\), the union of \(n\) consecutive 
intervals 
can become connected only if
\begin{equation}
    d<d_{n,\max},
\end{equation}
whereas an \(n\)-party HEGMEC exists only during time
intervals 
    $d_{n-1}(l,t)<d<d_n(l,t)$. The peak values of the critical curves therefore determine which collective 
    structures can ever become accessible in a given spatial configuration. Some fewer-party channels may 
    be activated transiently, while still shorter-range channels remain absent during
the entire quench.

To summarize, the two examples reveal a characteristic reorganization
of the connectivity hierarchy. During the growth stage, {the increasing correlation range} can temporarily 
open connectivity channels for the entanglement wedges of smaller unions of intervals. During the 
subsequent relaxation stage, these shorter-scale channels are screened and close successively. In this 
stage, a connected 
structure at fixed nearest-neighbor separation then requires an increasingly large collection of intervals 
and consequently spans a larger total spatial distance. In this precise sense, the characteristic support 
of the surviving collective correlations shifts from shorter-range to longer-range structures during 
thermalization.

The critical curves provide a more refined description of this process than \(I_n\) alone. The HEGMEC 
windows determine the minimal number of consecutive regions required to support a connected entanglement 
wedge at each stage of the evolution, and the values of \(I_n\) on HEGMEC configurations characterize the 
global entanglement 
shared by these $n$ parties. The observed transitions characterize how the spatial range, minimal 
collective support, and connectivity hierarchy of correlations evolve as the correlation front propagates 
and is subsequently modified by thermal screening.

\noindent{\bf III. The late-time plateau.}

For sufficiently large \(l\), the late-time critical separations approach a common thermal value,
\begin{equation}
    d_{2,f}\simeq d_{3,f}\simeq\cdots\simeq d_{n,f}
    \longrightarrow d_f=\frac{\log 2}{r_H}.
\end{equation}
Thus, the $n$-dependence of \(d_n\) curves is primarily a nonequilibrium feature of the intermediate 
thermalization regime and the initial state. This final saturation value is calculated as follows. 
In the final state, the hierarchy is compressed to a single screening scale determined by the temperature. 
{Meanwhile, since the AdS$_3$-Vaidya geometry describes a unitary evolution from an initially pure state, 
the full boundary state remains pure throughout the quench, and thermalization only occurs locally.} At 
sufficiently late times, the reduced state of any fixed finite region is described by the thermal state 
associated with the final BTZ geometry. Accordingly, once all HRT geodesics entering the calculation of 
$d_n$ lie entirely in the post-shell region, their lengths are given by
\begin{equation}
    L_{\mathrm{th}}(w)   =2\log\left[\frac{\sinh\left(\frac{r_H w}{2}\right)}{r_H}
    \right].
   \label{eq:Lthermal}
\end{equation}
% Thus, $d_{n,f}$ characterizes multipartite \zy{entanglement wedge?} connectivity in the locally
% thermal late-time state of the pure Vaidya evolution.\sun{should we delete this sentence?}

At the $n$-party transition, all proper subsets of the $n$ intervals are already in the disconnected phase. 
Their entropies therefore factorize into sums of single-interval entropies, and all fewer-partite 
information quantities vanish. $I_n$ then reduces to
\begin{equation}
    (-1)^n I_n
    =
    nS(l,t)-S_n(l;d,\ldots,d;t)
    =
    \frac{c_{\mathrm{CFT}}}{6}
    \left[
        nL(l,t)-\mathcal L_n^{\mathrm{con}}(l,d,t)
    \right],
    \label{eq:In_HEGMEC_reduction}
\end{equation}
where the fully connected candidate is
\begin{equation}
    \mathcal L_n^{\mathrm{con}}(l,d,t)
    =
    L\!\left(nl+(n-1)d,t\right)
    +(n-1)L(d,t).
    \label{eq:Ln_connected}
\end{equation}
The final critical separation is therefore determined exactly by the degeneracy between the fully connected 
and fully disconnected HRT configurations,
\begin{equation}
    nL_{\mathrm{th}}(l)
    =
    L_{\mathrm{th}}\!\left(nl+(n-1)d_{n,f}\right)
    +(n-1)L_{\mathrm{th}}(d_{n,f}).
    \label{eq:dnf_exact}
\end{equation}

For $r_Hl\gg1$, the geodesic lengths associated with $l$ and $nl+(n-1)d$ can be expanded as
\begin{equation}
    L_{\mathrm{th}}(w)
    =
    r_Hw-2\log(2r_H)
    +O(e^{-r_Hw}).
\end{equation}
The length $L_{\mathrm{th}}(d)$ must be kept exact because the critical separation might remain of order 
$r_H^{-1}$. Substituting these expressions into \eqref{eq:dnf_exact} gives
\begin{equation}
    r_Hd_{n,f}
    +
    2\log\sinh\left(\frac{r_Hd_{n,f}}{2}\right)
    =
    -2\log2,
\end{equation}
whose solution is
\begin{equation}
    d_{n,f}
    =
    \frac{\log2}{r_H}.
    \label{eq:dnf_asymptotic}
\end{equation}
The leading finite-$l$ correction is
\begin{equation}
    r_Hd_{n,f}
    =
    \log2
    -
    \frac{n}{2(n-1)}e^{-r_Hl}
    +O\!\left(e^{-2r_Hl},e^{-nr_Hl}\right).
    \label{eq:dnf_correction}
\end{equation}
Hence the independence of $d_{n,f}$ from $l$ and $n$ is asymptotic and not exact. For $r_H=1$, the
approximation is already accurate at the
percent level for $l\gtrsim5$, while at $l=10$ the different values of
$d_{n,f}$ are numerically indistinguishable and approach
$\log2\simeq0.693147$.

\begin{figure}[H]
  \centering
  \includegraphics[width=0.8\textwidth]{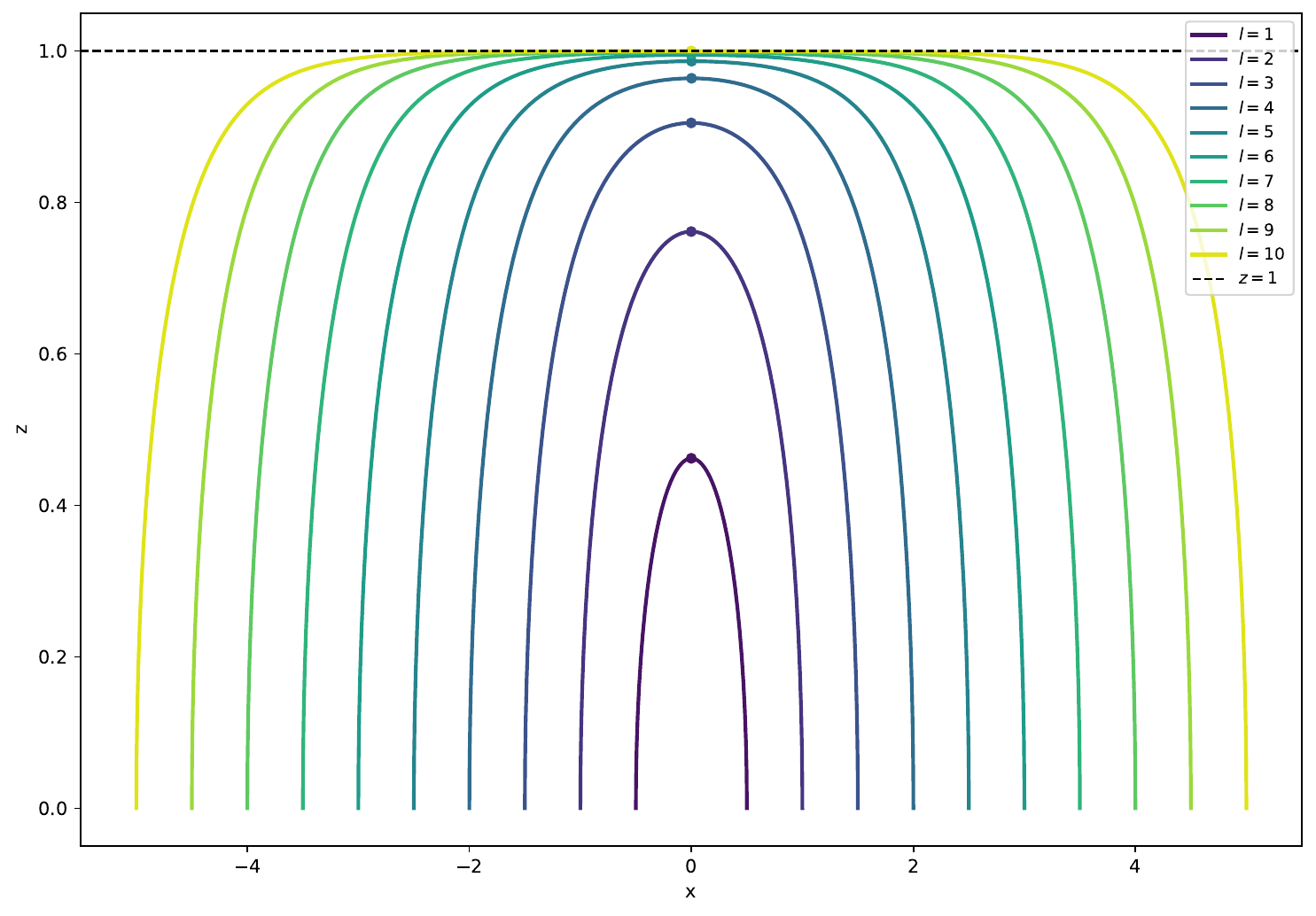}
  \caption{HRT surfaces in the BTZ spacetime for $r_H=1$ and $l=1,2,\ldots,10$.}
  \label{fig:geo}
\end{figure}

The large-$l$ universality of $d_{n,f}$ also admits a simple geometric
interpretation. In the late-time BTZ geometry, the equal-time geodesic
anchored on an interval of length $l$ remains outside the event horizon and
has turning point
\begin{equation}
    z_\star(l)
    =
    \frac{1}{r_H}
    \tanh\left(\frac{r_H l}{2}\right).
\end{equation}
Thus, as $l$ increases, the geodesic approaches the horizon at
$z_H=1/r_H$ without crossing it, as illustrated in
Figure~\ref{fig:geo} for $r_H=1$ and $l=1,\ldots,10$. In the large-interval
limit, the geodesic can be viewed approximately as two radial segments
extending from the boundary toward the horizon, joined by a long
near-horizon segment. When the fully connected and fully disconnected
$n$-interval configurations are compared, the extensive near-horizon
contributions proportional to the total interval length cancel. The
transition is therefore controlled only by the local geometry associated
with the separations between adjacent intervals, and this explains why the final
critical distance becomes independent of both $l$ and $n$.

\subsubsection{Integrated strength of the HEGMEC signal}

The critical separations \(d_n(l,t)\) characterize the spatial range
over which an \(n\)-region connected {entanglement wedge} can exist. However, they do not determine the
magnitude of the corresponding multipartite
information inside this range. Two configurations may have similar
values of \(d_n\), while carrying substantially different values of
\((-1)^n I_n\). To make the time dependence of the multipartite
information itself directly visible, we therefore introduce an
integrated measure of the HEGMEC signal.

Within an \(n\)-party HEGMEC, we define
\begin{equation}
    \mathcal{J}_n
    \left(
        A_{i_1}:\cdots:A_{i_n}
    \right)
    \equiv
    (-1)^n
    I_n
    \left(
        A_{i_1}:\cdots:A_{i_n}
    \right).
\end{equation}
For the holographic configurations considered here,
\(\mathcal{J}_n\) is nonnegative inside the corresponding HEGMEC
window. The integrated strength is then defined by
\begin{equation}
    E_n(t)
    =
    \int_{d_{n-1}(t)}^{d_n(t)}
    \sum_{1\leq i_1<\cdots<i_n\leq N}
    \mathcal{J}_n
    \left(
        A_{i_1}:\cdots:A_{i_n};t
    \right)
    d\mu(d),
    \label{eq:integrated-En}
\end{equation}
where \(N\) is the total number of intervals and $d\mu$ denotes the Lebesgue measure on $\mathbb{R}$. 
For the pairwise
case, we set \(d_1(t)=0\). Only configurations whose complete
\(n\)-region wedge is connected contribute {to \(E_n(t)\)} in the present symmetric
setup; the remaining terms vanish.

The quantity \(E_n(t)\) should be understood as the cumulative
HEGMEC signal accessible over the allowed separation window. It
combines two pieces of information: the width of the window in which
an exclusive \(n\)-region structure can occur and the magnitude of
\((-1)^n I_n\) inside that window. In this respect, \(E_n\) supplements
the critical curve \(d_n\): \(d_n\) determines where the multipartite entanglement structure is
allowed to exist, whereas \(E_n\) makes visible how much integrated
multipartite-information signal is supported within that range.

For the six-interval configuration studied here, the pairwise
integrated strength is
\begin{equation}
\begin{aligned}
    E_2(t)
    =
    \int_{0}^{d_2(t)}
    \Big[
        &5 I_2(l,d,t)
        +4 I_2(l,l+2d,t) \\
        &+3 I_2(l,2l+3d,t)
        +2 I_2(l,3l+4d,t) \\
        &+I_2(l,4l+5d,t)
    \Big]
    d\mu(d).
\end{aligned}
\end{equation}
The numerical coefficients count the number of interval pairs with
the same relative separation.

Similarly, the tripartite integrated strength is
\begin{equation}
\begin{aligned}
    E_3(t)
    =
    -\int_{d_2(t)}^{d_3(t)}
    \Big[
        &4 I_3(l,d,d,t)
        +6 I_3(l,d,l+2d,t) \\
        &+4 I_3(l,d,2l+3d,t)
        +2 I_3(l,d,3l+4d,t) \\
        &+2 I_3(l,l+2d,l+2d,t) \\
        &+2 I_3(l,l+2d,2l+3d,t)
    \Big]
    d\mu(d),
\end{aligned}
\end{equation}
where the overall minus sign follows from
\(\mathcal{J}_3=-I_3\geq0\) in the three-party HEGMEC regime.

\begin{figure}[H]
  \centering
  \includegraphics[width=0.75\textwidth]{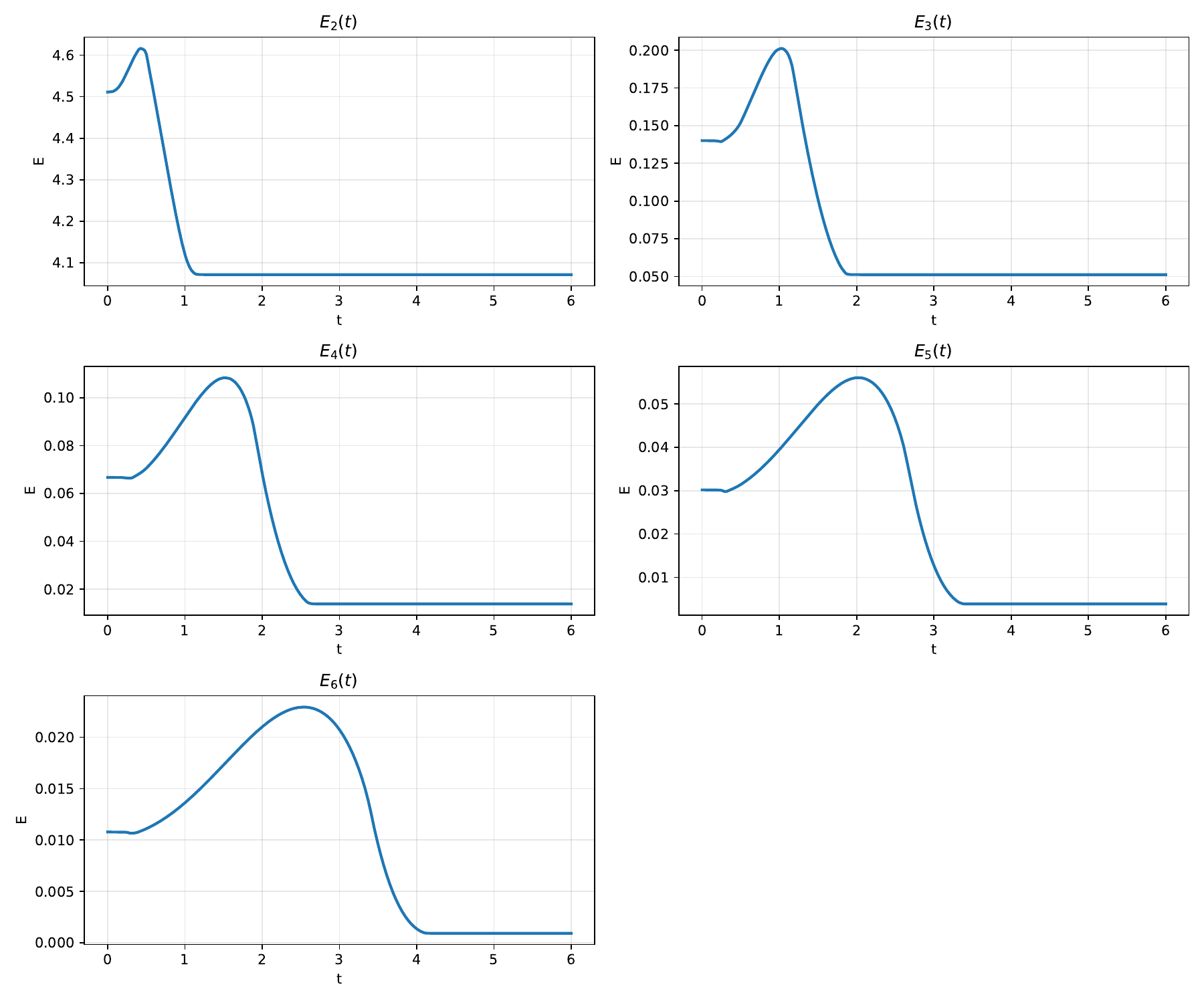}
  \caption{Time evolution of $E_n(t)$ for $r_H=1$ and $l=1$.}
  \label{fig:el1}
\end{figure}

\begin{figure}[H]
  \centering
  \includegraphics[width=0.75\textwidth]{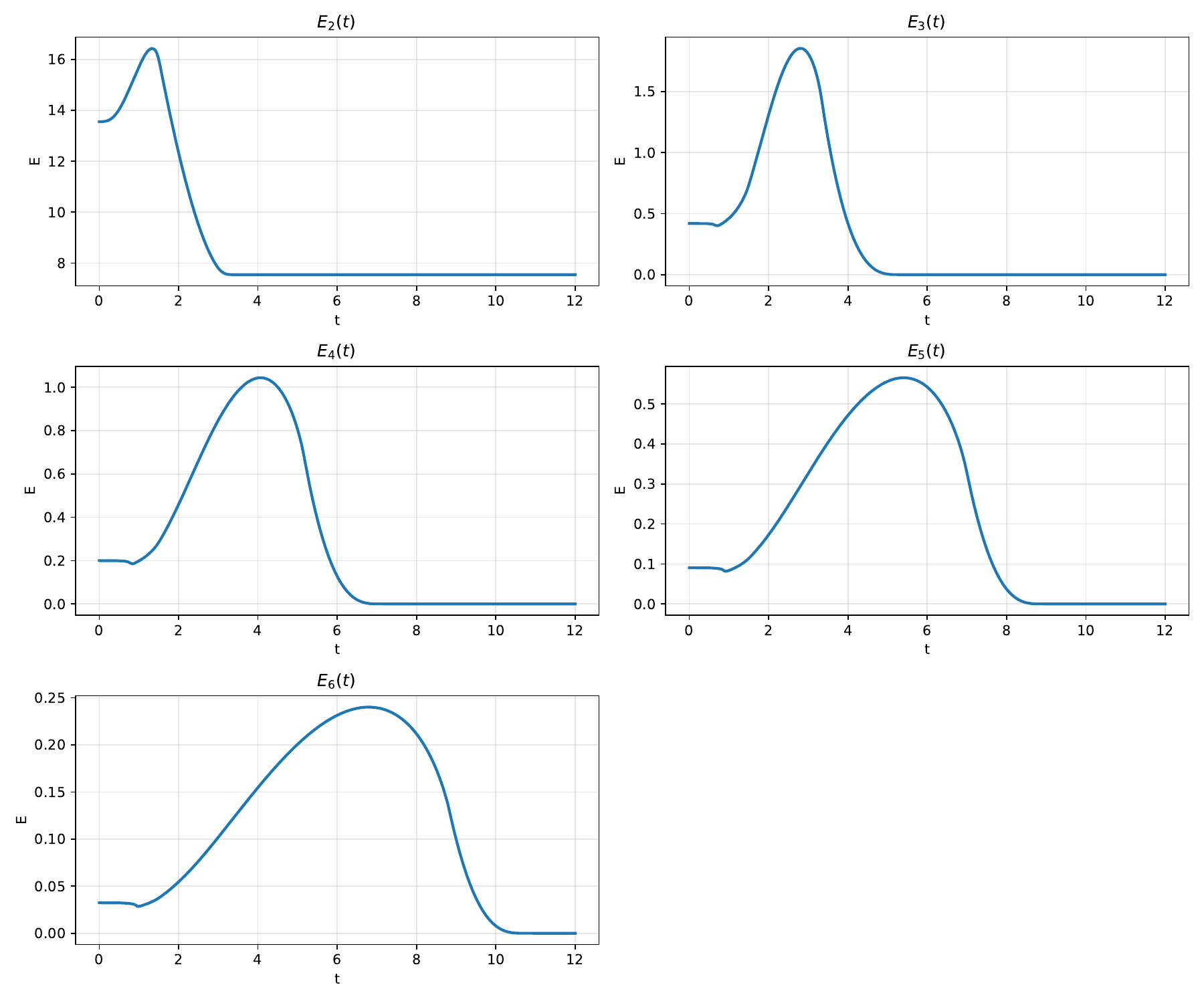}
  \caption{Time evolution of $E_n(t)$ for $r_H=1$ and $l=3$.}
  \label{fig:el3}
\end{figure}

\begin{figure}[H]
  \centering
  \includegraphics[width=0.75\textwidth]{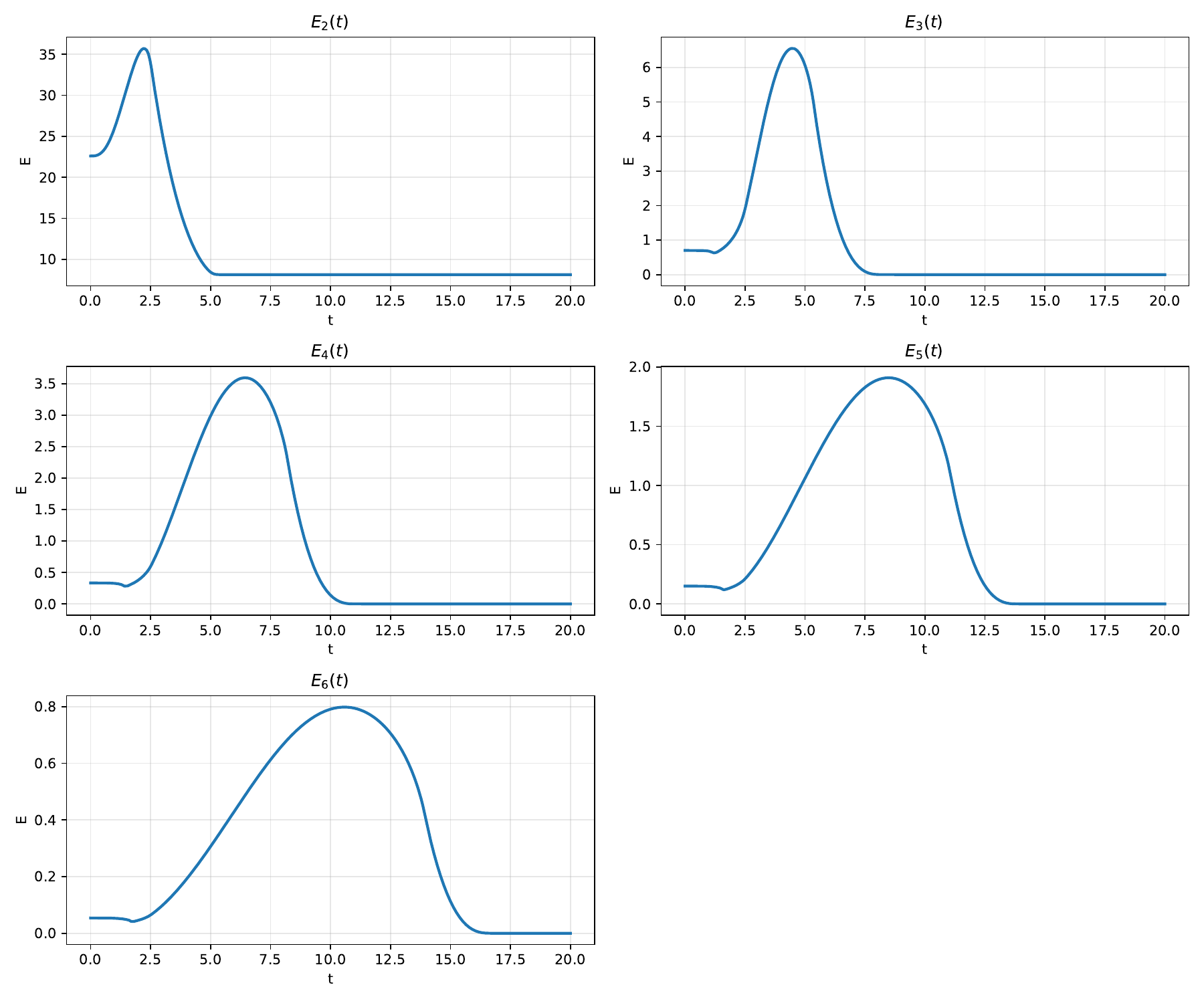}
  \caption{Time evolution of $E_n(t)$ for $r_H=1$ and $l=5$.}
  \label{fig:el5}
\end{figure}

For \(r_H=1\) and \(l=1,3,5\), the time dependence of \(E_n(t)\) is
shown in Figures~\ref{fig:el1}, \ref{fig:el3}, and
\ref{fig:el5}, which provide a direct visualization of the
growth and decay of the cumulative signal associated with each
order \(n\).


For all the interval lengths considered, the \(E_n(t)\) curves first
increase, reach a maximum, and subsequently decrease toward their
late-time values. The initial transient increase is produced by the quench in the cumulative HEGMEC window.
This increase
can result both from the widening of the allowed HEGMEC window and
from the growth of \((-1)^n I_n\) within that window. The subsequent
decrease indicates that the available separation range and/or the
signal value carried by the allowed configurations are reduced as the
system approaches the final locally thermal state.

% Thus, the nonmonotonic behavior of \(E_n\) contains information that
% cannot be extracted from the critical curves alone. An increasing
% \(d_n\) only shows that the corresponding connected structure can be
% supported across larger separations. It does not determine whether
% the multipartite-information signal is weak or strong within that
% range.\zy{Is this paragraph a little trivial? line 2 is sth already said and line 3 just restates the
%definition?} The simultaneous increase of \(d_n\) and \(E_n\) therefore
% indicates that the quench enlarges the accessible connectivity range
% while also increasing the cumulative signal supported by the
% associated configurations.

The peak times of the different integrated strengths are not
identical. For the numerical configurations studied here, they obey
the ordering
\begin{equation}
    t_{n,\max}^{(E)}
    >
    t_{n-1,\max}^{(E)}.
\end{equation}
More-party integrated signals therefore reach their maxima later
than fewer-party ones. This delayed peak cannot be interpreted
as a literal conversion of fewer-party entanglement into more-party entanglement, since the
quantities \(E_n\) are
constructed from different collections of boundary regions and integrated over different HEGMEC windows.
Instead, this
observed ordering shows that collective entanglement structures involving a larger number of subregions 
have a
delayed cumulative response to the quench and remain dynamically active on longer time scales.

The results therefore reveal two complementary aspects of the
thermalization process. The nearly coincident maxima of the critical
distances \(d_n(t)\) show that the spatial connectivity ranges of
different numbers of intervals expand on similar time scales, whereas the ordered
maxima of \(E_n(t)\) show that the accumulated multipartite-information
signals within these ranges develop and relax {in an $n$ dependent manner}. The geometric range and the
integrated
signal strength consequently encode distinct features of the
nonequilibrium entanglement structure.

\subsection{Upper bound of $-I_3$ and thermal entropy}

The preceding analysis characterized generic HEGMECs through the
critical separations \(d_n(t)\) and the integrated signals \(E_n(t)\).
We now turn to a more restrictive configuration introduced earlier:
the upper-bound configuration of the tripartite information.

For two fixed intervals \(a\) and \(b\) with $I(a:b)=0$, we
optimize a third region \(e\) and define {the upper bound of $-I_3$ as}
\begin{equation}
    \mathcal{U}_{\rho}(a,b)
    \equiv
    \sup_{e\subset\Sigma\setminus(a\cup b)}
    \left[-I_3(a:b:e)\right],
\end{equation}
where \(\Sigma\) denotes the physical boundary on which the state
\(\rho\) is defined. 
%The role of \(e\) resembles that of the separation \(d\) in the previous calculations in the broad sense 
%that 
%both are used to tune entanglement-wedge connectivity. Their optimization problems are nevertheless rather 
%different. In the symmetric HEGMEC configurations studied above, \(d\) is a single geometric parameter. By 
%contrast,\zy{just: rather different: the optimizing region $e$...} the optimizing region \(e\) in the 
%upper 
%bound configurations can contain many disconnected components, whose number, positions, and sizes must all 
%be adjusted.

{In \cite{Ju:2024kuc,Ju:2024hba}, the upper bound of $-I_3$ across a broad class of geometries was 
investigated systematically, including those where the information-theoretic upper bound could be 
saturated, as well as 
special geometries, such as wormholes, where the actual upper bound fails to reach the information-theoretic one. As a result, we have obtained a general upper
bound of $-I_3$ which is valid for arbitrary bulk geometries and dimensions. For the class of holographic 
configurations considered here, this general upper bound is:}
\begin{equation}
    \mathcal{U}_{\rho}(a,b)
    =
    \inf_{\substack{
        a\subset A,\; b\subset B\\
        A\cup B=\Sigma,\; A\cap B=\varnothing
    }}
    I_{\rho}(A:B),
    \label{eq:upper-bound-eosp}
\end{equation}
where \(A\) and \(B\) form an auxiliary bipartition of the physical
boundary and contain \(a\) and \(b\), respectively. This general expression
is defined as the state-constrained-purification form of the upper bound.

In the limiting configuration where the information-theoretic upper bound is saturated and
\(S(a)\leq S(b)\), the corresponding configuration satisfies
\begin{equation}
    I(a:b)=0,
    \qquad
    I(a:e)=0,
    \qquad
    I(a:be)=2S(a),
    \label{eq:upper-bound-saturation}
\end{equation} 
{and the maximum value $-I_3=2S(a)$.}
Again, while \(a\) has no bipartite correlation with either \(b\) or \(e\) separately, its correlation with 
the composite region \(b\cup e\) is maximal.  All degrees of freedom of \(a\) that contribute to its 
entropy are purified within \(b\cup e\). {This upper bound configuration is a special type of HEGMEC} and 
\(-I_3\) acquires 
the interpretation of a fully tripartite global entanglement measure.

In a time-dependent Vaidya geometry, an upper-bound configuration would therefore require \(e\) to be 
reoptimized at every boundary time: an optimal choice at one time does not necessarily remain optimal at a 
later time, because the relevant HRT surfaces and connectivity
conditions evolve during the quench. A complete covariant
optimization of \(e(t)\) is beyond the scope of the present work.
Instead, we compare the endpoint states in late-time Vaidya geometries with those in genuine thermal BTZ 
settings and focus on the effect of global purity on the optimized tripartite information.

\paragraph{Late-time pure Vaidya state.}

The state generated by the Vaidya quench remains globally pure under
unitary time evolution. At late times, however, the reduced density
matrix of any fixed finite interval becomes locally thermal, and its
entropy agrees with the corresponding BTZ result:
\begin{equation}
    S_{\mathrm{V}}(a,t\rightarrow\infty)
    =
    S_{\mathrm{BTZ}}(l),
\end{equation}
where \(l\) is the length of \(a\).

Global purity implies that, for every auxiliary bipartition
\(A\cup B=\Sigma\),
\begin{equation}
    S_{\mathrm{V}}(A)=S_{\mathrm{V}}(B),
    \qquad
    S_{\mathrm{V}}(\Sigma)=0,
\end{equation}
and hence
\begin{equation}
    I_{\mathrm{V}}(A:B)=2S_{\mathrm{V}}(A).
\end{equation}
For the symmetric configuration considered here, the infimum is
obtained by shrinking the auxiliary region \(A\) to the fixed
interval \(a\). The late-time upper bound is therefore
\begin{equation}
    \mathcal{U}_{\mathrm{V}}(l)
    =2S(a)=
    2S_{\mathrm{BTZ}}(l).
    \label{eq:pure-Vaidya-upper-bound}
\end{equation}

This result is important conceptually. Although the entropy of $a$ has the same numerical value as in a 
thermal state, it does not represent the entanglement between $a$ and a thermal purifier outside the 
system. Instead, 
since the global state is pure, this entropy is entirely generated by correlations between $a$ and the 
remaining degrees of freedom within the same boundary CFT.
By optimizing \(e\), these purifying degrees of freedom in
principle can be incorporated into \(b\cup e\), allowing the bound
\(2S(a)\) to be reached.

\paragraph{Thermal BTZ state.}

We next compare the above calculation with the corresponding result for the thermal BTZ state, restricting 
the optimizing region $e$ to the same physical boundary, where the configuration of $a$ and $b$ is kept 
unchanged. The thermal state is mixed and may be regarded
as part of a purification
\begin{equation}
    |\Psi\rangle_{\Sigma Q},
\end{equation}
where \(Q\) denotes an inaccessible purifying system. In a
thermofield-double representation, \(Q\) is the second asymptotic
boundary.

{While the local entropy \(S(a)\) is the same in the two states at late times, the distinction between this 
genuine thermal BTZ state and the pure Vaidya state lies in the location of the degrees of freedom that 
purify \(a\)}: the region \(e\subset\Sigma\) cannot access
the purifier \(Q\).
Consequently, not all of the entropy of \(a\) can necessarily be
organized into tripartite entanglement among \(a\), \(b\), and \(e\).

To evaluate the resulting correction, we set \(r_H=1\) and introduce
a large regulator interval \(\Sigma_{L_0}\) of length \(L_0\). Let
\(A\) be a central interval of length \(L_A\), and let
\begin{equation}
    B=\Sigma_{L_0}\setminus A
\end{equation}
be its two-component complement inside the regulator interval. We work in renormalized geodesic-length 
units, suppressing the
overall factor \(c/6\). The thermal geodesic length of an interval of
length \(w\) is
\begin{equation}
    L_{\mathrm{th}}(w)
    =
    2\ln\left[\sinh\left(\frac{w}{2}\right)\right].
\end{equation}
The entropy of \(B\) is determined by two competing RT
configurations:
\begin{equation}
    L(B)
    =
    \min\left\{
        L_{\mathrm{th}}(L_0)+L_{\mathrm{th}}(L_A),
        \;
        2L_{\mathrm{th}}
        \left(\frac{L_0-L_A}{2}\right)
    \right\}.
    \label{eq:B-two-branches}
\end{equation}
The first candidate contains the surface homologous to the large
regulator interval together with the surface of \(A\), whereas the
second candidate consists of two disconnected surfaces associated
with the two components of \(B\).

It follows that
\begin{equation}
\begin{aligned}
    I_{\mathrm{th}}(A:B)
    &=
    L_{\mathrm{th}}(L_A)
    +L(B)
    -L_{\mathrm{th}}(L_0).
\end{aligned}
\end{equation}
Taking the limit \(L_0\rightarrow\infty\), and suppressing the common
UV-divergent term, gives
\begin{equation}
    I_{\mathrm{th}}(A:B)
    =
    \min\left\{
        2L_{\mathrm{th}}(L_A),
        \;
        L_{\mathrm{th}}(L_A)-L_A-2\ln 2
    \right\}.
    \label{eq:thermal-auxiliary-MI}
\end{equation}

Both branches in  \eqref{eq:thermal-auxiliary-MI} increase
monotonically with \(L_A\). The infimum compatible with
\(a\subset A\) is therefore obtained for the smallest allowed
auxiliary region,
\begin{equation}
    L_A=l.
\end{equation}
The upper bound restricted to the physical thermal boundary is thus
\begin{equation}
    \mathcal{U}_{\mathrm{th}}(l)
    =
    \min\left\{
        2L_{\mathrm{th}}(l),
        \;
        L_{\mathrm{th}}(l)-l-2\ln 2
    \right\}.
    \label{eq:thermal-upper-bound}
\end{equation}

\paragraph{Finite upper-bound deficit.}

We define
\begin{equation}
\Delta\mathcal{U}(l)\equiv \mathcal{U}_{\mathrm{V}}(l)-  \mathcal{U}_{\mathrm{th}}(l).
\end{equation}
{Note that this difference is finite and independent of the common UV regulator.} Using \eqref{eq:pure-Vaidya-upper-bound} and
\eqref{eq:thermal-upper-bound}, we obtain
\begin{align}
    \Delta\mathcal{U}(l)
    =
    \max\left\{
        0,
        \;
        L_{\mathrm{th}}(l)+l+2\ln 2
    \right\}.
    \label{eq:upper-bound-deficit}
\end{align}

\begin{figure}[H]
    \centering
    \includegraphics[width=0.7\textwidth]{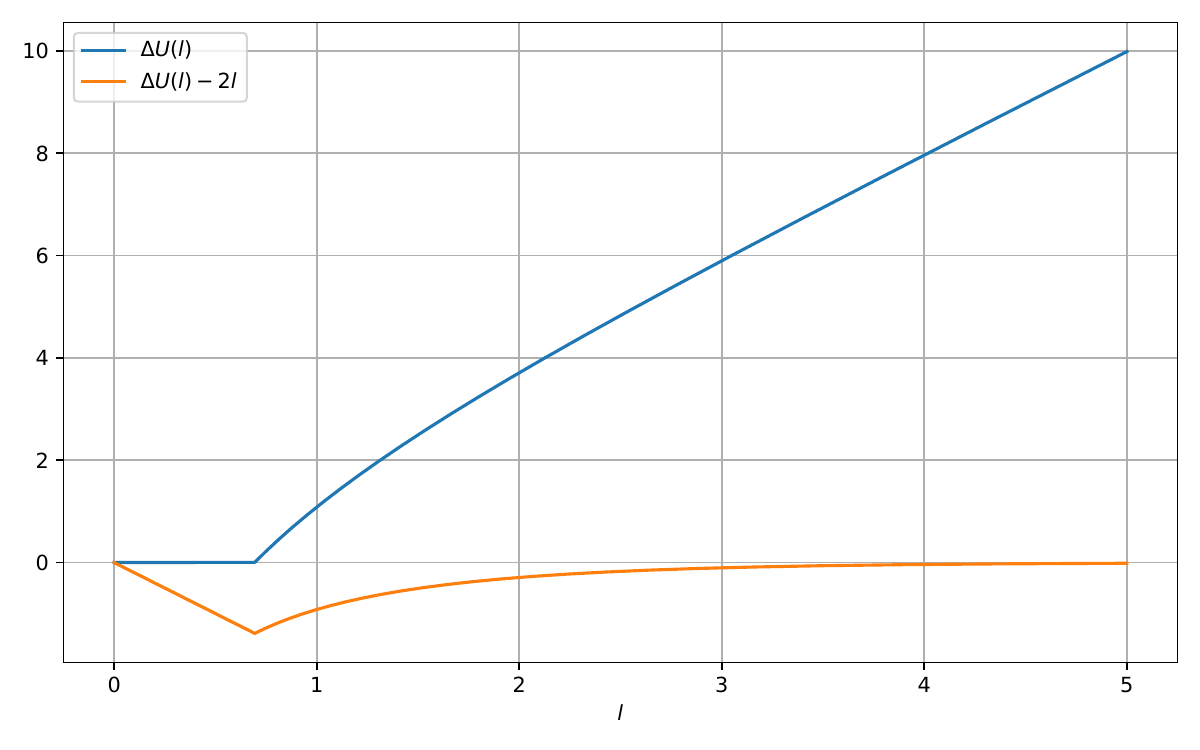}
    \caption{
        The finite upper-bound deficit
        \(\Delta\mathcal{U}(l)\) and the deviation
        \(\Delta\mathcal{U}(l)-2l\) from its large-\(l\)
        volume-law behavior.
    }
    \label{fig:diff}
\end{figure}

Equivalently,
{\begin{equation}
    \Delta\mathcal{U}(l)
    =
    \begin{cases}
        0,
        & l\leq \ln 2,\\[4pt]
        2\ln\left(e^l-1\right),
        & l>\ln 2.
    \end{cases}
    \label{eq:upper-bound-deficit-piecewise}
\end{equation}}
The behavior of $\Delta\mathcal{U}(l)$ is shown in Figure \ref{fig:diff}.

For intervals shorter than the critical scale $l<l_c=\ln 2$, the first RT branch in  \eqref{eq:B-two-branches} 
dominates. In
this branch, the extensive thermal contribution contained in the
large-\(L_0\) surface cancels in the mutual information
\(I(A:B)\). Consequently,
\begin{equation}
    \mathcal{U}_{\mathrm{th}}(l)
    =
    \mathcal{U}_{\mathrm{V}}(l),
\end{equation}
and the optimized upper bound of $-I_3$ cannot distinguish the thermal
mixed state from the late-time pure Vaidya state.

This means that in the thermal mixed state, subsystems smaller than this critical length are not 
bipartitely 
entangled with the external purifier $Q$. Nevertheless, such subsystems still possess thermal corrections 
to the values of their entanglement entropy. Their full correlation budget is therefore exhausted by 
correlations with the complementary region \(B\) on the same physical boundary. The critical length \(l_c\) 
marks the onset 
at which the subsystem becomes  sufficiently large to develop leading-order correlations with the external 
purifier {$Q$}. For a generic finite-temperature quantum state, the entropy \(S(A)\) cannot in general be 
uniquely decomposed into a contribution from correlations within the physical system and a contribution 
associated with the external thermal purifier. However, here in this holographic setup, the  optimized 
upper bound of \(-I_3\) gives a direct diagnostic of how much of the correlation budget of a subsystem $a$ 
remains accessible within the physical boundary and how much is occupied by the inaccessible thermal 
purifier. 

For \(l>l_c\), the disconnected RT configuration for \(B\) becomes
dominant. The thermal contribution no longer cancels completely, and
the upper bound in the mixed state becomes smaller than that in the
globally pure state. The deficit \(\Delta\mathcal{U}\) measures the
part of the entropy of \(a\) which is purified by the external purifier $Q$ and cannot be purified by
regions \(b\) and \(e\) on the physical boundary alone.

At large \(l\),
\begin{equation}
\begin{aligned}
    \Delta\mathcal{U}(l)
    &=
    2\ln\left(e^l-1\right)\\
    &=
    2l-2e^{-l}+O(e^{-2l}).
\end{aligned}
\end{equation}
Restoring the factor \(c/6\), the leading behavior is
\begin{equation}
    \Delta\mathcal{U}_{S}(l)
    \simeq
    \frac{c}{3}l
    =
    2s_{\mathrm{th}}l,
\end{equation}
where \(s_{\mathrm{th}}=c/6\) for \(r_H=1\). The upper-bound deficit
therefore reproduces the volume-law scaling of the thermal entropy in the large subsystem limit.
The factor of two originates from the information-theoretic bound
\(2S(a)\).  In the large-\(l\) limit, an extensive volume-law part of the entropy is asymptotically 
associated with the external purifier. Correlations with the physical complement remain, but they are 
nonextensive and are dominated by contributions localized near the entangling surface.

{Note that for \(l\leq l_c=\ln 2/r_H\), the vanishing upper-bound deficit $\Delta\mathcal{U}(l)=0$ admits 
another direct purification-theoretic interpretation.
Recall that \(Q\) is the external system that purifies the thermal state on the physical boundary \(A\cup 
B\). 
Since the complete state on
\(ABQ\) is pure, one has
\begin{equation}
    I(A:B)+I(A:Q)=2S(A).
\end{equation}
In the present configuration,
\(\mathcal{U}_{\mathrm{th}}=I(A:B)\), whereas the pure-state upper
bound is \(2S(A)\). Therefore,
\begin{equation}
    \Delta\mathcal{U}=I(A:Q).
\end{equation}

The vanishing of \(\Delta\mathcal{U}\) for \(l\leq l_c\) thus means
that, at leading order in the classical holographic limit, the
interval \(A\) has no mutual information with the external thermal
purifier. Its full correlation budget \(2S(A)\) is exhausted by
correlations with the complement \(B\) on the same physical
boundary. Consequently, the restricted optimization over
\(e\subset B\) can reproduce the pure-state upper bound.

For \(l>l_c\), one instead finds
\begin{equation}
    I(A:Q)=\Delta\mathcal{U}>0.
\end{equation}
A finite part of the correlation budget of \(A\) is then carried by
the inaccessible purifier \(Q\). Since the optimizing region \(e\)
is restricted to the physical boundary, these purifying degrees of
freedom cannot participate in the tripartite structure among
\(a\), \(b\), and \(e\). The maximal value of \(-I_3\) is therefore
reduced relative to the globally pure state.}

Interestingly, the critical length at which the external purifier
becomes visible coincides with the universal late-time separation,
\begin{equation}
    l_c=d_f=\frac{\ln 2}{r_H}.
\end{equation}
This coincidence reflects a common reorganization of correlation
channels induced by the BTZ horizon. The scale \(d_f\) is the maximal
nearest-neighbor separation across which large subregions on the same
physical boundary can maintain a connected entanglement wedge,
whereas \(l_c\) is the minimal subsystem size at which leading-order
correlations with the external purifier become nonzero. The former
therefore characterizes thermal screening of same-boundary
connectivity, while the latter characterizes the onset of
cross-purifier correlations. Both scales originate from competing connected and disconnected RT
configurations in the BTZ geometry.

In summary, the comparison in this subsection provides information complementary to the preceding
HEGMEC analysis. The quantities \(d_n(t)\) and \(E_n(t)\) describe
how the range and integrated strength of generic collective
correlations evolve during unitary thermalization. The upper-bound
deficit instead probes the global purification structure of the
state. The late-time Vaidya state and the thermal BTZ state have the
same entanglement entropies for fixed finite intervals, but they have
different optimized tripartite capacities because one is globally
pure whereas the other contains inaccessible correlations with an
external purifier. The upper-bound construction therefore distinguishes local
thermalization from genuine global thermal mixedness.

\section{Markov gap in AdS$_3$-Vaidya}\label{sec4}

In this section, we study another tripartite measure: the Markov gap in the AdS\(_3\)-Vaidya geometry.
Consider two disjoint boundary intervals \(A\) and \(B\). Their Markov gap is defined as
\begin{equation}
    h(A:B)
    =
    S_R(A:B)-I(A:B),
    \label{eq:markov-gap-definition}
\end{equation}
where $I(A:B)$ is the mutual information and \(S_R(A:B)\) is the reflected entropy
\cite{Dutta:2019gen,Hayden:2021gno}.

To see the physical meaning of the Markov gap, consider the
globally pure tripartite state \(|\psi\rangle_{ABC}\), where
\(C=\overline{A\cup B}\). A useful reference entanglement structure is given by the triangle state 
\cite{Zou:2020bly}, for which the local Hilbert spaces can be factorized
as
\begin{equation}
    \mathcal{H}_{\alpha}
    =
    \mathcal{H}_{\alpha_L}\otimes\mathcal{H}_{\alpha_R},
    \qquad \alpha=A,B,C,
\end{equation}
and
\begin{equation}
    |\psi\rangle_{ABC}
    =
    |\psi\rangle_{A_LB_R}
    \otimes
    |\psi\rangle_{B_LC_R}
    \otimes
    |\psi\rangle_{A_RC_L}
\end{equation}
up to local unitary transformations.
Thus, after decomposing each subsystem into smaller factors, the
entanglement is entirely organized into the three pairs of bipartite correlations.
% \zy{, while the genuine irreducible tripartite entanglement among all three subsystems is absent.} 
%\sun{in 
%the following sentences, we say that there is tripartite entanglement among the coarse grained subsystems 
%abc?}
Importantly, such a state may still be genuinely tripartite entangled according to the usual definition 
based on the coarse subsystems \(A\), \(B\), and \(C\); what is absent is the irreducible tripartite 
entanglement at the 
level of the finer subsystem decomposition.

In \cite{Zou:2020bly}, a vanishing Markov gap has been shown to reveal that the tripartite pure state 
belongs 
to the class of sums of triangle states (SOTS), in which different
triangle-state sectors are combined through an appropriate direct-sum
decomposition of the local Hilbert spaces. Triangle states form a
special subclass of SOTS, while other non-triangle states such as the GHZ state can also be
SOTS. Consequently,
\begin{equation}
    h(A:B)>0
\end{equation}
signals a non-SOTS-type tripartite entanglement structure, which
cannot be reduced to a collection of pairwise entangled links even after subdividing the local Hilbert 
spaces. 
Conversely,
\(h(A:B)=0\) does not imply the absence of tripartite entanglement
in the standard coarse-grained definition.

In holography, the Markov gap and the related quantity
\(g(A:B)=2E_P(A:B)-I(A:B)\) have the same leading semiclassical
geometric dual,
\begin{equation}
    h(A:B)
    =
    2E_W(A:B)-I(A:B),
\end{equation} where \(E_W(A:B)\) is the entanglement wedge cross
section (EWCS) \cite{Takayanagi:2017knl,Nguyen:2017yqw} and \(E_P(A:B)\) is the entanglement of purification
\cite{Terhal:2002riz}. The holographic Markov gap therefore diagnoses whether the
entanglement among \(A\), \(B\), and \(C\) can be organized into a
triangle- or SOTS-like network after the subsystems are resolved into
smaller factors \cite{Bao:2025psl}. This complements the HEGMEC analysis of the previous
section: while \(d_n(t)\) and \(E_n(t)\) characterize the range and strength
of collective correlations among $n$-party HEGMECs, the Markov gap probes the internal structural class
of the tripartite entanglement of the whole boundary state during thermalization.

However, \(h(A:B)\) singles out the pair \(A,B\), and is not manifestly symmetric under permutations of the 
tripartition \(A,B,C\), which violates the property required for a tripartite
entanglement measure. Following
 \cite{Chen:2026xtx}, we define the permutation-symmetric
tripartite Markov gap by
\begin{equation}
    h_3(A:B:C)
    =
    \min\left\{
        h(A:B),\,
        h(B:C),\,
        h(A:C)
    \right\}.
    \label{eq:tripartite-markov-gap}
\end{equation}
This minimum prescription is particularly useful for asymmetric
tripartitions. In the present setup, $C=\overline{A\cup B}$ is a noncompact region, so \(h(A:C)\) and
\(h(B:C)\) generally contain divergent contributions. A geometric mean of the
three pairwise Markov gaps would therefore be ill-defined \cite{Basak:2023uix,Chen:2026xtx}. Meanwhile,
 \eqref{eq:tripartite-markov-gap} remains finite and selects the
weakest tripartite channel:
% For the configurations studied below, \(h(A:B)\) is finite, while the
% two quantities involving the noncompact complement \(C\) are
% divergent. 
\begin{equation}
    h_3(A:B:C)=h(A:B).
\end{equation}
Thus, although we calculate the Markov gap from the finite pair
\(A,B\), the resulting quantity should be understood as the
permutation-symmetric tripartite Markov gap of the full pure-state
partition \(A\cup B\cup C\).

\subsection*{Results in the Vaidya geometry}

Now we calculate the time dependence of the Markov gap in the Vaidya geometry. Consider two boundary 
intervals
\(A\) and \(B\), each of length \(l\)
and separated by a distance \(d\). The Markov gap is only nonzero when the entanglement wedge of $AB$
is connected. The mutual information is given by
\begin{equation}
    I(A:B)
    =
    2S(l,t)-S(d,t)-S(2l+d,t),
    \label{eq:MI-two-intervals}
\end{equation}
and the Markov gap is obtained from
\begin{equation}
    h(l,d,t)
    =
    2E_W(l,d,t)-I(A:B).
    \label{eq:markov-gap-Vaidya}
\end{equation}

To evaluate the EWCS, let
\begin{equation}
    (z_d,v_d)
    =
    \bigl(z_p(d,t),v_p(d,t)\bigr),
    \qquad
    (z_u,v_u)
    =
    \bigl(z_p(2l+d,t),v_p(2l+d,t)\bigr)
\end{equation}
denote the turning points of the HRT surfaces associated with the
inner interval of width \(d\) and the outer interval of width
\(2l+d\), respectively. The two characteristic saturation times are
\begin{equation}
    t_d=\frac{d}{2},
    \qquad
    t_u=\frac{2l+d}{2}.
\end{equation}
The EWCS then has three branches \cite{BabaeiVelni:2020wfl},
\begin{equation}
E_W(l,d,t)
=
\begin{cases}
\displaystyle
\ln\left[
    \frac{z_u}{z_d}
    \frac{
        1+\sqrt{1+Q_a^2z_u^2}
    }{
        1+\sqrt{1+Q_a^2z_d^2}
    }
\right],
& 0<t<t_d,
\\[5mm]
\displaystyle
\ln\left[
    \frac{z_u}{z_d}
    \frac{
        1+\sqrt{1+Q_a^2z_0^2}
    }{
        1+\sqrt{1+Q_a^2z_u^2}
    }
    \frac{
        z_H+\sqrt{z_H^2+(Q_b^2-1)z_d^2}
    }{
        z_H+\sqrt{z_H^2+(Q_b^2-1)z_0^2}
    }
\right],
& t_d<t<t_u,
\\[5mm]
\displaystyle
\ln\left[
    \frac{z_u}{z_d}
    \frac{
        z_H+
        \sqrt{z_H^2+(Q_b^2-1)z_u^2}
    }{
        z_H+
        \sqrt{z_H^2+(Q_b^2-1)z_d^2}
    }
\right],
& t>t_u,
\end{cases}
\label{eq:EWCS-Vaidya-piecewise}
\end{equation}
where \(z_H=1/r_H\), and \(z_0\) is the radial coordinate at which
the EWCS intersects the null shell \(v=0\). In the first regime both
HRT surfaces cross the shell; in the intermediate regime only the
outer HRT surface crosses it; and in the final regime both surfaces
lie entirely in the BTZ region.

The parameters \(Q_a\) and \(Q_b\) are fixed by the matching
conditions across the shell:
\begin{equation}
    Q_a^2
    =
    \frac{
        4(v_u+z_u-z_0)^2
    }{
        v_u(v_u-2z_0)
        (v_u+2z_u)
        (v_u+2z_u-2z_0)
    },
    \label{eq:Qa}
\end{equation}
and
\begin{equation}
    \frac{
        Q_bz_H-
        \sqrt{z_H^2+(Q_b^2-1)z_d^2}
    }{
        Q_bz_H-
        \sqrt{z_H^2+(Q_b^2-1)z_0^2}
    }
    \frac{z_H+z_0}{z_H+z_d}
    =
    e^{v_d/z_H}.
    \label{eq:Qb}
\end{equation}

For an interval of width \(w\), the turning-point coordinates {of its HRT surface} are
\begin{equation}
    z_p(w,t)
    =
    \begin{cases}
    \displaystyle
    \frac{1}{r_\star(w,t)},
    & 0<t<\dfrac{w}{2},
    \\[3mm]
    \displaystyle
    z_H\tanh\left(\frac{r_Hw}{2}\right),
    & t\geq\dfrac{w}{2},
    \end{cases}
    \label{eq:zp-width}
\end{equation}
and
\begin{equation}
    v_p(w,t)
    =
    \begin{cases}
    \displaystyle
    \frac{1}{r_c(t)}
    -
    \frac{1}{r_\star(w,t)},
    & 0<t<\dfrac{w}{2},
    \\[3mm]
    \displaystyle
    t+\dfrac{1}{2r_H}
    \log\left[
    \dfrac{1-\tanh\left(\frac{r_H w}{2}\right)}
          {1+\tanh\left(\frac{r_H w}{2}\right)}
  \right],
    & t\geq\dfrac{w}{2}.
    \end{cases}
    \label{eq:vp-width}
\end{equation}
Here \(r_\star(w,t)\) is the turning radius of the HRT surface and
\(r_c(w,t)\) is its actual shell-crossing radius. They are obtained
from equations~\eqref{eq:r_lambda}-\eqref{eq:x_lambda} and
\eqref{eq:x_in}.

\begin{figure}[H]
    \centering
    \includegraphics[width=\textwidth]{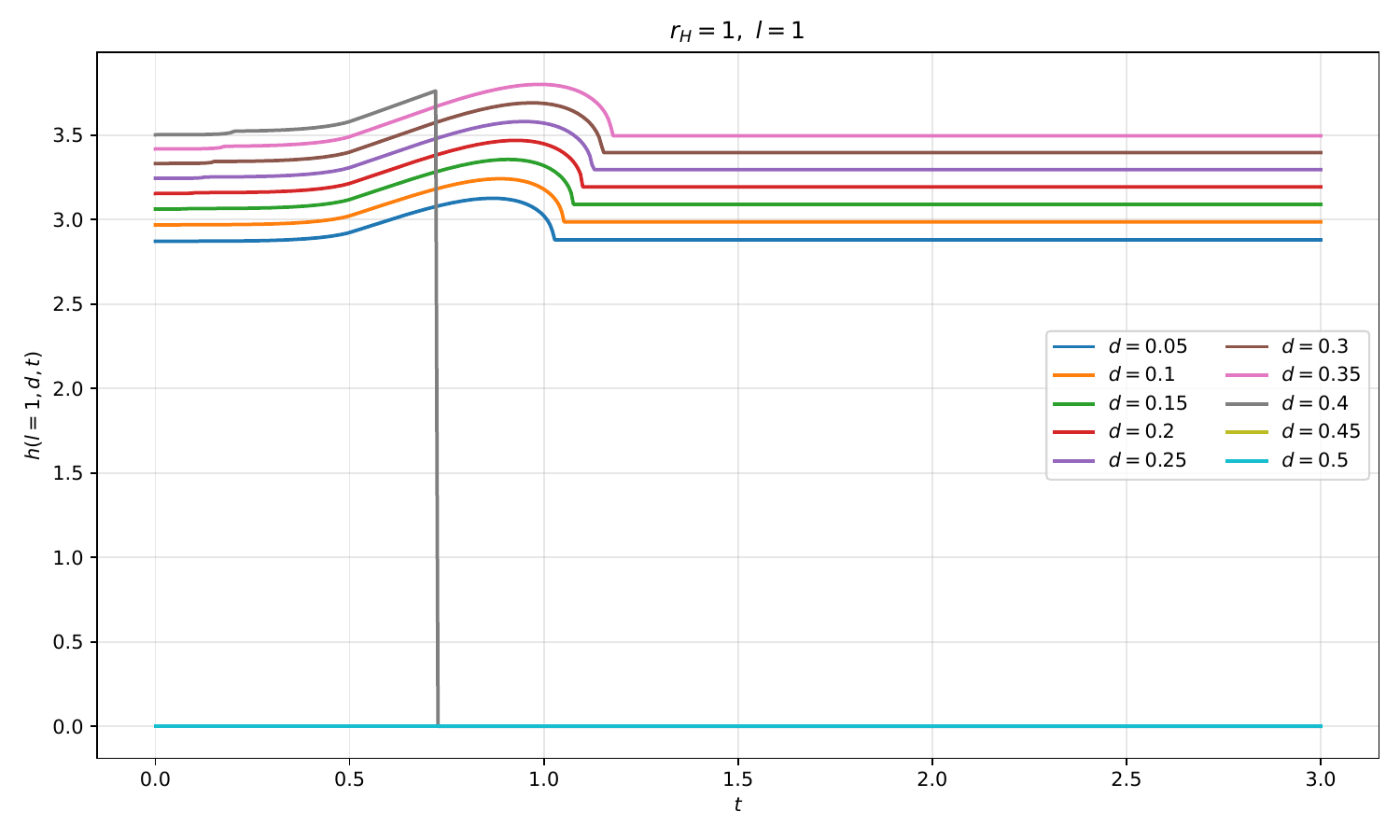}
    \caption{
        Time evolution of the Markov gap for \(r_H=1\), \(l=1\), and
        \(d=0.05,0.10,\ldots,0.50\). The Markov gap is zero when the entanglement wedge of \(AB\) is
        disconnected.
        The abrupt drop for \(d=0.40\) marks the
        connected-to-disconnected entanglement-wedge transition,
        whereas the configurations with \(d=0.45\) and \(0.50\)
        remain disconnected throughout the evolution.
    }
    \label{fig:markov-gap}
\end{figure}

For each \(l\), \(d\), and \(t\), we first determine whether the
entanglement wedge is connected. In the connected phase, we evaluate
the appropriate branch of  \eqref{eq:EWCS-Vaidya-piecewise},
combine it with the mutual information \eqref{eq:MI-two-intervals}, and obtain the Markov
gap from  \eqref{eq:markov-gap-Vaidya}. As an example, we set
\(r_H=1\), \(l=1\), and
\begin{equation}
    d=0.05,0.10,\ldots,0.50.
\end{equation}
The resulting time evolution is shown in
Figure~\ref{fig:markov-gap}.

The time dependence of the Markov gap is controlled by the critical
separation \(d_2(l,t)\) introduced in the previous section. For a fixed
configuration, the entanglement wedge of \(AB\) is connected when $d<d_2(l,t)$ and disconnected when 
\(d>d_2(l,t)\). Figure~\ref{fig:markov-gap}
therefore exhibits three distinct regimes.

\noindent{\bf Regime I:}

For sufficiently small separations, represented by
\(d=0.05,\ldots,0.35\), the entanglement wedge remains connected
throughout the thermalization process. The Markov gap is consequently
positive at all times. It first increases, reaches a maximum during the
intermediate nonequilibrium stage, and then decreases to a constant
late-time value. The final plateau is slightly higher than the initial
vacuum value. In terms of the interpretation introduced above, the initial growth indicates a transient 
increase in the collective entanglement shared among \(A\), \(B\), and their complement beyond the 
description of SOTS correlations. The later decrease represents a partial {suppression} of this non-SOTS-type tripartite structure {as the system relaxes}. Nevertheless, because the final value remains above the 
vacuum value, the 
late-time locally thermal state retains a stronger Markov-gap signal for these fixed subregions than the 
initial state. 

The fact that the late-time plateau lies above the vacuum value indicates that the locally thermalized 
state contains a stronger non-SOTS-type tripartite entanglement structure for the fixed regions \(A\), 
\(B\), and 
{their complement} \(C\). This is consistent with the redistribution of entanglement toward larger 
collective spatial scales revealed in the
previous section, since shorter-scale correlations within each subregion are naturally more compatible with 
a triangle- or SOTS-like decomposition  after the local Hilbert spaces are further factorized. As 
thermalization 
reorganizes part of this structure
into correlations extending over larger spatial scales, it becomes more difficult to transform the state 
into a 
triangle or SOTS form by local unitary transformations acting separately within \(A\), \(B\), and \(C\). 
 
\noindent{\bf Regime II:}

An intermediate behavior is illustrated by \(d=0.40\). In this case, the entanglement wedge is initially 
connected, but as \(d_2(l,t)\) decreases during the relaxation stage, it eventually crosses the fixed value 
\(d=0.40\). At this time, the dominant HRT configuration for \(S(AB)\) changes from the connected saddle to 
the disconnected one, which leads to
\begin{equation}
    I(A:B)=E_W(A:B)=h(A:B)=0.
\end{equation}
The discontinuous drop of the Markov gap is expected: the mutual
information approaches zero continuously at the transition, whereas
the EWCS generally remains finite on the connected side and vanishes
once the entanglement wedge becomes disconnected.

\noindent{\bf Regime III:}

The larger separations \(d=0.45\) and \(0.50\) remain above $d_2(l,t)$
throughout the entire evolution. In these cases, the entanglement wedge of $AB$ is always disconnected, and 
the corresponding Markov gaps remain zero. Note that by definition, this only indicates that the particular 
Markov-gap channel between $A$ and $B$ is inaccessible for the fixed spatial configurations, while global 
tripartite entanglement may still exist.

More generally, the nonmonotonic evolution of \(d_2(l,t)\) leads to a re-entrant dynamics
\begin{equation}
    h(A:B)=0
    \quad\longrightarrow\quad
    h(A:B)>0
    \quad\longrightarrow\quad
    h(A:B)=0
\end{equation}
for any fixed separation satisfying
\begin{equation}
    d_2(l,0)<d<d_{2,\max}(l).
\end{equation}
In such a configuration, the quench temporarily connects an initially disconnected entanglement wedge, and 
generates a detectable non-SOTS-type tripartite entanglement structure, which is subsequently removed by 
thermal screening. 

We finally emphasize an important distinction between the present
analysis and the pure-thermal comparison performed for the
tripartite information in the previous section. The Vaidya quench
produces a globally pure boundary state. Consequently, $C=\overline{A\cup B}$
is the complete purifier of \(\rho_{AB}\), and the full state
\(|\psi\rangle_{ABC}\) admits the triangle-state and SOTS
interpretations discussed above. In particular, the thermal-form
entropy of the late-time finite regions is purified by degrees of
freedom contained in \(C\) on the same boundary.

By contrast, a genuine single-boundary thermal BTZ state is globally
mixed. The physical complement \(C\) does not by itself purify
\(\rho_{AB}\); a complete purification requires an additional system
\(Q\),
\begin{equation}
    |\Psi\rangle_{ABCQ},
\end{equation}
which may be represented by the second boundary of the
thermofield-double state. Although the bipartite quantity
\(h(A:B)=S_R(A:B)-I(A:B)\) can still be defined for the thermal mixed state, it no longer has the same 
interpretation as a tripartite triangle/SOTS diagnostic of the physical partition \(A|B|C\).
A direct comparison between the pure Vaidya result and a single-boundary thermal BTZ result would therefore 
compare different purification structures. For this reason, unlike the upper-bound analysis of \(I_3\), we 
do not use the Markov gap to make a direct pure-state versus thermal mixed-state comparison. Thus, the late-time Markov gap studied here characterizes the multipartite organization of a pure state that is locally 
thermal, 
rather than the entanglement structure of a globally thermal mixed state.

\section{Genuine tripartite multi-entropy in AdS$_3$-Vaidya}\label{sec5}

{In this section, we turn to a complementary quantity\textemdash the genuine tripartite multi-entropy, 
which is 
constructed from the multi-entropy \cite{Gadde:2022cqi} to directly isolate genuine tripartite entanglement 
\cite{Iizuka:2025ioc,Iizuka:2025caq}.} 

{The multi-entropy $S^{(q)}(A_1:\dots:A_q)$ extends the notion of entanglement entropy to multipartite 
quantum systems, and captures the correlation shared among the $q$ systems $A_1,\dots, A_q$. It is defined 
by analytically continuing the replica parameter of Rényi multi-entropies to one, while the latter is 
constructed through higher-dimensional toric-lattice contractions of replicated density matrix tensors.
When $q=2$, the multi-entropy reduces to the usual von Neumann entropy:}
\begin{equation}
    S^{(2)}(A_1:A_2)
    =
    S(A_1)
    =
    S(A_2).
\end{equation}

{The multi-entropy $S^{(q)}$ generally contains both genuine $q$-partite entanglement and fewer-party 
correlations \cite{Gadde:2023zzj}. When $q=3$, in order to subtract the latter while isolating the genuine 
tripartite contribution shared by all three parties, the genuine multi-entropy $GM^{(3)}$ has been 
introduced in 
\cite{Iizuka:2025ioc,Iizuka:2025caq} as:
\begin{equation}
\begin{aligned}
    GM^{(3)}(A:B:C)
    &=
    S^{(3)}(A:B:C)
    \\
    &\quad
    -\frac{1}{2}
    \left[
        S^{(2)}(A:BC)
        +S^{(2)}(B:CA)
        +S^{(2)}(C:AB)
    \right]
    \\
    &=
    S^{(3)}(A:B:C)
    -\frac{1}{2}
    \left[
        S(A)+S(B)+S(C)
    \right].
    \label{eq:genuine-tri-entropy}
\end{aligned}
\end{equation}
}
In particular, \(GM^{(3)}\) vanishes for separable states
and triangle states, whose entanglement can be resolved into
independent bipartite links after an appropriate factorization of the local Hilbert spaces. A nonzero 
\(GM^{(3)}\) therefore signals the tripartite entanglement beyond the triangle-state structure.

{We then consider the holographic prescription for the multi-entropy. In a static geometry,} the multi-entropy is conjectured to be given by the area of a minimal soap-film network \(W_{\min}\) in the bulk 
\cite{Gadde:2022cqi} that partitions the bulk Cauchy slice into chambers homologous to the corresponding 
boundary subsystems:
\begin{equation}
    S^{(q)}(A_1:\cdots:A_q)
    =
    \frac{\operatorname{Area}(W_{\min})}{4G_N}.
    \label{eq:holo-multi-static}
\end{equation}
When several admissible topologies exist, the one with the smallest total area is selected.

For time-dependent geometries, \cite{Gadde:2022cqi} proposed a covariant extension in which the static 
minimal network is replaced by an extremal Lorentzian soap-film configuration. In the following, we apply 
this proposed 
prescription to the AdS\(_3\)-Vaidya geometry and study the evolution of the genuine tripartite multi-entropy during thermalization.

We consider the following boundary tripartition:
\begin{equation}
    A=[-l,0],
    \qquad
    B=[0,l],
    \qquad
    C=\overline{A\cup B}
    =
    (-\infty,-l)\cup(l,\infty).
    \label{eq:tri-partition}
\end{equation}
Thus, \(A\) and \(B\) are two adjacent intervals of equal length \(l\), while \(C\) is their complement 
with 
\(ABC\) being a pure tripartite system. For this partition, the bulk soap film consists of three geodesic 
segments connecting a bulk junction point \(p\) to the three boundary interfaces
\begin{equation}
    x=-l,\qquad x=0,\qquad x=l
\end{equation}
at the same boundary time \(t\). We therefore first summarize the geodesic building blocks required for the 
calculation.

A geodesic segment lying entirely in the post-shell region \(v>0\) is described by \eqref{eq:r_lambda}-
\eqref{eq:x_lambda}, with its conserved parameters \(E\) and \(J\) fixed by the two endpoints. In the pre-shell 
pure-AdS region \(v<0\), it is convenient to use the Poincar\'e time coordinate \(\tau=v+z\). For two bulk 
endpoints \((\tau_1,z_1,x_1)\) and \((\tau_2,z_2,x_2)\), the geodesic can be parametrized as
\begin{align}
    z(\lambda)
    &=
    \frac{
        z_1z_2\sinh s
    }{
        z_2\sinh(s-\lambda)+z_1\sinh\lambda
    },
    \label{eq:z_ads}
    \\
    x(\lambda)
    &=
    \frac{
        z_2x_1\sinh(s-\lambda)
        +z_1x_2\sinh\lambda
    }{
        z_2\sinh(s-\lambda)+z_1\sinh\lambda
    },
    \label{eq:x_ads}
    \\
    \tau(\lambda)
    &=
    \frac{
        z_2\tau_1\sinh(s-\lambda)
        +z_1\tau_2\sinh\lambda
    }{
        z_2\sinh(s-\lambda)+z_1\sinh\lambda
    },
    \label{eq:t_ads}
\end{align}
where \(0\leq\lambda\leq s\), with the two endpoints located at \(\lambda=0\) and \(\lambda=s\).

Let the bulk junction be
\begin{equation}
    p=(v_p,z_p,x_p),
\end{equation}
and consider a geodesic connecting \(p\) to a boundary point \((v=t,z=0,x)\). If \(p\) lies in the \(v<0\) 
region, the geodesic crosses the null shell once and is composed of a pure-AdS segment and a BTZ segment. 
If \(p\) lies in the \(v>0\) region, the geodesic may either remain entirely in the BTZ region or enter the 
pure-AdS region and cross the shell twice. At every shell crossing, the two segments are joined by imposing 
continuity and the corresponding refraction conditions. When several admissible branches exist, the one 
with the smallest renormalized length is selected.

Denote the renormalized length of the geodesic connecting \(p\) to the boundary point \((t,x)\) by
\begin{equation}
    L_R(p;x,t)
    \equiv
    L_R(v_p,z_p,x_p;x,t).
\end{equation}
The renormalized length of the minimal three-leg soap film is then
\begin{equation}
\begin{aligned}
    \Gamma(t)
    =
    \min_{\text{admissible branches}}
    \operatorname*{ext}_{\substack{
        z_p>0\\
        v_p,x_p\in\mathbb{R}
    }}
    \Big[
        &L_R(p;-l,t)
        +L_R(p;0,t)
        +L_R(p;l,t)
    \Big].
    \label{eq:Gamma-soap-film}
\end{aligned}
\end{equation}
The extremization over \(p\) determines the trivalent junction and implements the local force-balance 
condition among the three geodesic legs. If several extremal networks exist, the one with the smallest 
total renormalized 
length gives the multi-entropy,
\begin{equation}
    S^{(3)}(A:B:C;t)
    =
    \frac{\Gamma(t)}{4G_N}.
\end{equation}

The bipartite terms entering the genuine tripartite multi-entropy are computed from the HRT surfaces. For 
the present symmetric configuration,
\begin{equation}
    S(A)=S(B)=\frac{L(l,t)}{4G_N},
\end{equation}
whereas global purity gives $S(C)=S(A\cup B)=\frac{L(2l,t)}{4G_N}$. Here \(L(w,t)\) denotes the 
renormalized 
HRT length of an interval of width \(w\), as defined in  \eqref{eq:vaidya_length}. The genuine tripartite 
multi-entropy is therefore
\begin{equation}
    GM^{(3)}(A:B:C;t)
    =
    \frac{1}{4G_N}
    \left[
        \Gamma(t)
        -
        \frac{1}{2}
        \left(
            2L(l,t)+L(2l,t)
        \right)
    \right].
    \label{eq:GM3-Vaidya}
\end{equation}
Each of the three boundary interfaces contributes the same universal UV divergence to the soap-film network 
and to the subtracted HRT terms. These divergences cancel exactly in
 \eqref{eq:GM3-Vaidya}, leaving a finite result.

As a consistency check, before the quench the geometry is pure AdS\(_3\). The extremal soap film then 
reduces 
to the hyperbolic
Steiner tree, whose three geodesic legs meet at mutual angles of
\(120^\circ\) \cite{Harper:2024ker}. Substituting its length into
 \eqref{eq:GM3-Vaidya} gives
\begin{equation}
    GM_{\mathrm{vac}}^{(3)}(A:B:C)
    =
    \frac{c}{2}
    \log\frac{2}{\sqrt{3}},
    \label{eq:vac-genuine-tri}
\end{equation}
which is independent of \(l\) and provides the initial value of the time-dependent result.

\begin{figure}[H]
    \centering
    \includegraphics[width=0.8\textwidth]{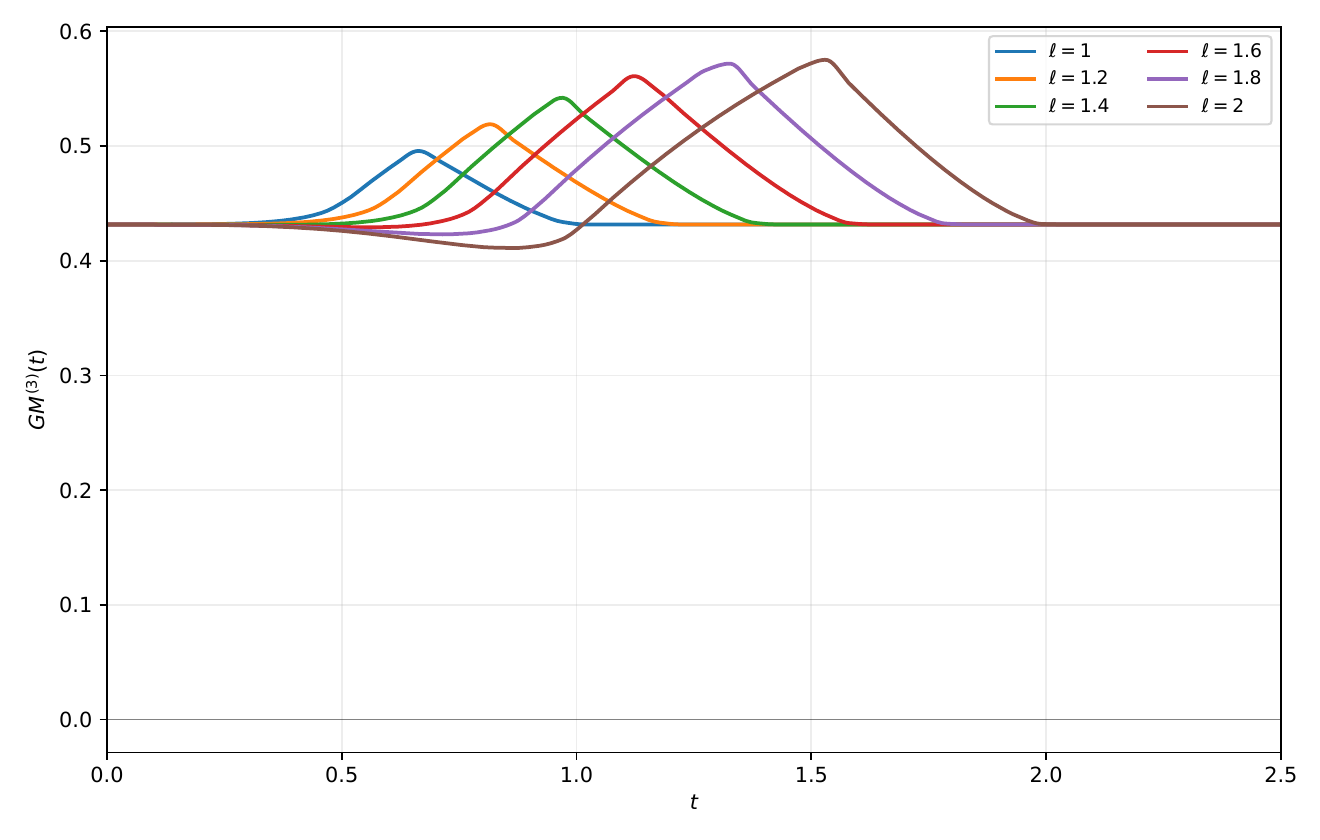}
    \caption{
        Time evolution of the genuine tripartite multi-entropy for two adjacent
        intervals with \(r_H=1\) and
        \(l=1.0,1.2,\ldots,2.0\).
    }
    \label{fig:GM3}
\end{figure}

For the numerical calculation, we set \(r_H=1\) and take
\begin{equation}
    l=1.0,1.2,\ldots,2.0.
\end{equation}
The resulting time evolution of the genuine tripartite multi-entropy is shown in
Figure~\ref{fig:GM3}.

The time evolution shown in Figure~\ref{fig:GM3} is generally
nonmonotonic. For the smaller intervals, the genuine tripartite multi-entropy first increases and then 
decreases, whereas for sufficiently large \(l\) it exhibits an additional initial suppression before its 
enhancement. This behavior follows from the fact that \(GM^{(3)}\) is a difference between the three-leg 
soap-film length and the bipartite HRT contributions. Its time derivative is therefore
\begin{equation}
    \frac{dGM^{(3)}}{dt}
    =
    \frac{1}{4G_N}
    \left[
        \dot{\Gamma}(t)
        -\dot L(l,t)
        -\frac{1}{2}\dot L(2l,t)
    \right].
\end{equation}

The constituent geometric quantities respond to the quench on different time scales. The HRT surfaces 
associated with the intervals of widths \(l\) and \(2l\) saturate at
\begin{equation}
    t_{\mathrm{sat}}(l)=\frac{l}{2},
    \qquad
    t_{\mathrm{sat}}(2l)=l,
\end{equation}
respectively, while the trivalent network has its own delayed response. At early times, the bipartite HRT 
contributions can grow faster than the soap-film length, producing the initial decrease visible for the 
larger 
values of \(l\). Once the width-\(l\) HRT surfaces have saturated, \(\dot L(l,t)=0\), whereas the trivalent 
network continues to evolve. The genuine tripartite multi-entropy consequently begins to increase and 
reaches 
its maximum at a time satisfying approximately
\begin{equation}
    \dot{\Gamma}(t_{\mathrm{peak}})
    =
    \frac{1}{2}\dot L(2l,t_{\mathrm{peak}}).
\end{equation}
The peaks occur between \(l/2\) and \(l\), and move to later times as \(l\) increases. This shows that 
genuine 
tripartite structures supported on larger spatial scales are established later during the quench. The full 
observable settles to its final value around \(t\simeq l\), controlled by the largest finite scale \(2l\) 
in 
the partition.

The initial decrease of \(GM^{(3)}\) for sufficiently large \(l\) also admits a direct physical 
interpretation. 
Immediately after the quench, correlations are generated locally near the three entangling points and 
propagate 
outward. At this early stage, the newly generated entanglement is predominantly organized across individual 
cuts and is therefore closer to a collection of pairwise, triangle-state-like links. These bipartite 
contributions increase before the causal regions emitted from the different interfaces have sufficiently 
overlapped to establish a collective correlation involving \(A\), \(B\), and \(C\) simultaneously.
Consequently, the fraction of the entanglement detected as genuine tripartite structure initially 
decreases, 
even though the total entanglement of the state is increasing.

At intermediate times, the correlation fronts associated with the different interfaces begin to overlap. 
The 
entanglement can then no longer be organized solely into independent pairwise links, and a stronger 
irreducible 
three-region structure develops. This produces the subsequent growth and transient maximum of \(GM^{(3)}\). 
The initial suppression is more visible for larger \(l\), because the spatially separated fronts require a 
longer time to overlap; for smaller \(l\), the pairwise and tripartite buildup occur within a shorter 
common 
time window, so that only the growth and relaxation are clearly resolved.

At late times, after the relevant finite length scales have locally thermalized, the additional 
nonequilibrium 
three-region structure relaxes. The genuine tripartite multi-entropy therefore returns to its universal 
static 
value. A notable feature of Figure~\ref{fig:GM3} is that the late-time value returns to the vacuum result,
\begin{equation}
    GM_{\mathrm{final}}^{(3)}
    =
    GM_{\mathrm{vac}}^{(3)}
    =
    \frac{c}{2}\log\frac{2}{\sqrt{3}}.
\end{equation}
{Thus, the initial and final states have the same value of $GM^{(3)}(A:B:C)$, while their complete 
entanglement 
structures are not necessarily identical.}

This equality can be understood geometrically. On a static locally AdS\(_3\) time slice, the non-winding Y-shaped Steiner network anchored at three boundary points satisfies
\begin{equation}
    \Gamma_{\mathrm{static}}
    =
    \frac{1}{2}
    \left(
        L_{12}+L_{23}+L_{31}
    \right)
    +
    3\log\frac{2}{\sqrt{3}}.
\end{equation}
For the endpoints \(-l,0,l\), this becomes
\begin{equation}
    \Gamma_{\mathrm{static}}
    -
    \frac{1}{2}
    \left[
        2L(l)+L(2l)
    \right]
    =
    3\log\frac{2}{\sqrt{3}}.
\end{equation}
Both the initial AdS geometry and the late-time BTZ exterior admit such static constant-curvature slices. 
Consequently, once all relevant extremal surfaces lie in the final BTZ region and remain on the same 
topological branch, the same universal Steiner excess is recovered.

\section{Conclusion and outlook}\label{sec6}

{In this work, we studied the evolution of multipartite entanglement after a global quench in AdS\(_3\)-Vaidya. 
By restricting to HEGMEC configurations, we identified regimes in which \((-1)^n I_n\) isolates the 
collective 
entanglement shared by all \(n\) parties. The spatial range and integrated signal of this multipartite 
entanglement are transiently enhanced and subsequently reduced during thermalization. For intervals with 
fixed 
lengths and separations, during the time evolution, the surviving connected {entanglement wedge} structure 
requires increasingly large collections of subregions and therefore spans larger total spatial scales. This 
provides a multipartite refinement of the holographic picture in which entanglement propagates from shorter 
to longer distances.}

{At late times, however, same-boundary multipartite connectivity is restricted by the thermal scale 
\(d_f\): separations larger than \(d_f\) cannot support a leading-order connected entanglement wedge, 
irrespective of 
the number of participating regions. By comparing the upper-bound configurations in the late-time pure 
Vaidya state and the thermal BTZ state, we further identify a critical subsystem size \(l_c\), above which 
leading-order correlations with the external thermal purifier become visible. Remarkably, \(l_c=d_f\), 
indicating that 
the same thermal scale controls both the loss of long-range same-boundary connectivity and the onset of 
access 
to the thermal purifier. The Markov gap behavior during thermalization shows that the vanishing of 
collective 
entanglement in $n$-HEGMEC subregions in the final state does not necessarily imply a simpler entanglement 
structure. {Within the short-range channels of connected bipartite entanglement wedges}, the tripartite 
entanglement among two regions and their complement can become less reducible to a network of pairwise 
links. 
The genuine tripartite multi-entropy provides a complementary result: for the adjacent tripartition studied 
here, it undergoes a nontrivial intermediate evolution but returns to its vacuum value at
late times.}

These results provide a geometric description of how shorter-range entanglement evolves into longer-range 
entanglement during the holographic thermalization. An important question, however, remains open. The 
present 
analysis determines the spatial range of connected structures and the values of several multipartite 
signals, 
but it does not determine if any fewer-partite entanglement could also convert into more-partite 
entanglement 
during the thermalization process, or whether the contrary could happen. More refined spatial partitions or 
more multipartite entanglement measures would be needed to make it possible to test whether thermalization 
merely shifts the spatial support of correlations or also changes the microscopic proportion of few-partite 
and 
multipartite entanglement carried by each region.

\section*{Acknowledgement}
\noindent We thank Bo-Hao Liu for useful discussions. XXJ would like to acknowledge the support of the
Shuimu Tsinghua Scholar Program of Tsinghua University. 
This work was supported by Project  12575068 supported by the National Natural Science Foundation of China.

\appendix
\section{HRT geodesics in the AdS$_3$-Vaidya geometry}\label{appA}

Following the analytic thin-shell geodesic solutions of \cite{Balasubramanian:2011ur}, in this appendix we 
collect the formulas used in the calculations of Sections \ref{sec3}, \ref{sec4}, and \ref{sec5}. In 
particular, we review the general BTZ geodesic segments and the shell-crossing data used to determine the 
turning point and the renormalized length of a boundary-anchored HRT geodesic. The pure-AdS branch is 
standard 
and will not be reviewed separately.

\subsection{BTZ geodesic segments}

{A spacelike geodesic in the BTZ geometry can be parametrized by the proper length $\lambda$ as 
$g_{\mu\nu}\frac{dX^\mu}{d\lambda}\frac{dX^\nu}{d\lambda}=1$. It carries the dimensionless conserved 
quantities}
\begin{equation}
    E
    =
    \frac{r^2-r_H^2}{r_H}
    \frac{d\tau_{\mathrm B}}{d\lambda},
    \qquad
    J
    =
    \frac{r^2}{r_H}
    \frac{dx}{d\lambda}.
    \label{eq:EJ_definition}
\end{equation}
Here $E$ characterizes the displacement of the geodesic in the static time direction, while $J$ is the 
conserved momentum along the boundary spatial direction.

A branch approaching the right endpoint $(t_b,\ell/2)$ as $\lambda\to+\infty$ can be written as 
\cite{Balasubramanian:2011ur}
\begin{align}
r(\lambda)
&=
\frac {r_H} {2}
\sqrt{
\left[
e^{\lambda}
+\left((J+1)^2-E^2\right)e^{-\lambda}
\right]
\left[
e^{\lambda}
+\left((J-1)^2-E^2\right)e^{-\lambda}
\right]
},
\label{eq:r_lambda}
\\
\tau_{\mathrm B}(\lambda)
&=
t_b+
\frac{1}{2r_H}
\log
\left|
\frac{
e^{\lambda}
+
\left[J^2-(1+E)^2\right]e^{-\lambda}
}{
e^{\lambda}
+
\left[J^2-(1-E)^2\right]e^{-\lambda}
}
\right|,
\label{eq:t_lambda}
\\
x(\lambda)
&=
\frac{\ell}{2}
+
\frac{1}{2r_H}
\log
\left[
\frac{
e^{\lambda}
+
\left((J-1)^2-E^2\right)e^{-\lambda}
}{
e^{\lambda}
+
\left((J+1)^2-E^2\right)e^{-\lambda}
}
\right].
\label{eq:x_lambda}
\end{align}
Note that the time in \eqref{eq:t_lambda} is the static BTZ time $\tau_{\mathrm B}$, rather than the 
ingoing 
coordinate $v$.

For a complete equal-time boundary-anchored geodesic lying in the BTZ region, time-reflection
symmetry gives
\begin{equation}
    E=0,
    \qquad
    J=\coth\left(\frac{r_H\ell}{2}\right),
\end{equation}
and hence the corresponding renormalized geodesic length is 
\begin{equation}
    L_{\mathrm{BTZ}}^{\mathrm{ren}}(\ell)
    =
    2\log\left[
      \frac{\sinh\left(r_H\ell/2\right)}{r_H}
    \right].
    \label{eq:L_BTZ_ren}
\end{equation}

\subsection{Boundary-anchored geodesics crossing the null shell}

For $0<t_b<\frac{\ell}{2}$, the symmetric HRT geodesic crosses the shell at
\begin{equation}
    Q_\pm=(v=0,r=r_c,x=\pm x_c).
\end{equation}
The shell-crossing radius $r_c$ is determined by the boundary time,
\begin{equation}
r_c=r_c(t_b)=r_H\coth(r_H t_b).
\end{equation}

The central portion of the geodesic lies in pure AdS and reaches its radial
turning point $r=r_\star$ at $x=0$. Reflection symmetry implies that this
central segment lies on a constant Poincar\'e-time slice. Since $v=0$ at
$r=r_c$,  \eqref{eq:ads_EF_static} gives
\begin{equation}
    \tau_{\mathrm A}=\frac{1}{r_c}.
\end{equation}
The spatial profile of the AdS segment is
\begin{equation}
    x(r)
    =
    \frac{\sqrt{r^2-r_\star^2}}{r_\star r},
    \qquad
    r_\star\leq r\leq r_c,
    \label{eq:x_in}
\end{equation}
and hence
\begin{equation}
    x_c
    =
    \frac{\sqrt{r_c^2-r_\star^2}}{r_\star r_c}.
\end{equation}

The outer portions are BTZ geodesic segments of the form
\eqref{eq:r_lambda}-\eqref{eq:x_lambda}. Their constants $E$ and $J$ are
determined by matching the geodesic across the null shell. The position of
the curve is continuous at $v=0$, and the variation of the total length
imposes the corresponding refraction condition on its tangent momentum.
The resulting constants are
\begin{equation}
    E
    =
    -\frac{r_H\sqrt{r_c^2-r_\star^2}}{2r_c^2},
    \qquad
    J=\frac{r_\star}{r_H}.
    \label{eq:EJ_crossing}
\end{equation}
Unlike the equilibrium BTZ geodesic, the crossing segments have $E\neq0$:
although the two boundary endpoints occur at equal boundary time, each BTZ
segment connects a boundary point to a shell point at a different static
time.

For numerical evaluation, it is convenient to introduce the dimensionless
variables
\begin{equation}
    s=\frac{r_\star}{r_c},
    \qquad
    \chi=\sqrt{1-s^2},
    \qquad
    \rho=\frac{r_c}{r_H}.
    \label{eq:s_chi_rho}
\end{equation}
In these variables,
\begin{equation}
    E=-\frac{\chi}{2\rho},
    \qquad
    J=s\rho.
\end{equation}
The boundary-time condition determines the crossing radius:
\begin{equation}
    2\rho
    =
    \coth(r_Ht_b)
    +
    \sqrt{
       \coth^2(r_Ht_b)
       -
       \frac{2\chi}{1+\chi}
    }.
    \label{eq:rho_st}
\end{equation}
The remaining parameter $s=s(\ell,t_b)$ is determined implicitly by the
boundary separation:
\begin{equation}
    \ell
    =
    \frac{1}{r_H}
    \left[
       \frac{2\chi}{s\rho}
       +
       \log
       \left(
       \frac{
          2(1+\chi)\rho^2+2s\rho-\chi
       }{
          2(1+\chi)\rho^2-2s\rho-\chi
       }
       \right)
    \right].
    \label{eq:s_implicit}
\end{equation}
For fixed $(\ell,t_b)$ in the crossing regime, we solve
 \eqref{eq:s_implicit} for $s\in(0,1)$, use
 \eqref{eq:rho_st} to obtain $r_c=r_H\rho$, and then evaluate the
on-shell geodesic length. This converts the original extremal-curve problem
into a one-dimensional implicit root-finding problem.

Combining the geodesic solutions of these two regimes, together with the pure AdS solution, the renormalized
HRT geodesic lengths are
\begin{equation}
    L^{\mathrm{ren}}(\ell,t_b)
    =
    \begin{cases}
    2\log\left(\dfrac{\ell}{2}\right),
    & t_b\leq0,
    \\[4mm]
    2\log\left[
       \dfrac{\sinh(r_Ht_b)}
       {r_Hs(\ell,t_b)}
    \right],
    & 0<t_b<\dfrac{\ell}{2},
    \\[5mm]
    2\log\left[
       \dfrac{
          \sinh\left(\frac{r_H\ell}{2}\right)
       }{r_H}
    \right],
    & t_b\geq\dfrac{\ell}{2},
    \end{cases}
\end{equation}
as shown in \eqref{eq:vaidya_length} in the main text.
\bibliographystyle{JHEP}
\bibliography{ref}

@article{Erdmenger:2017gdk,
    author = "Erdmenger, Johanna and Fernandez, Daniel and Flory, Mario and Megias, Eugenio and Straub, Ann-Kathrin and Witkowski, Piotr",
    title = "{Time evolution of entanglement for holographic steady state formation}",
    eprint = "1705.04696",
    archivePrefix = "arXiv",
    primaryClass = "hep-th",
    reportNumber = "MPP-2017-87",
    doi = "10.1007/JHEP10(2017)034",
    journal = "JHEP",
    volume = "10",
    pages = "034",
    year = "2017"
}

@article{Maulik:2025blg,
    author = "Maulik, Sabyasachi and Pari, Soumen",
    title = "{Entanglement entropy and its linear response following a global quench in holographic Gauss-Bonnet gravity}",
    eprint = "2503.18090",
    archivePrefix = "arXiv",
    primaryClass = "hep-th",
    doi = "10.1016/j.nuclphysb.2025.117084",
    journal = "Nucl. Phys. B",
    volume = "1018",
    pages = "117084",
    year = "2025"
}

@article{Dur:2000zz,
    author = "Dur, W. and Vidal, G. and Cirac, J. I.",
    title = "{Three qubits can be entangled in two inequivalent ways}",
    eprint = "quant-ph/0005115",
    archivePrefix = "arXiv",
    doi = "10.1103/PhysRevA.62.062314",
    journal = "Phys. Rev. A",
    volume = "62",
    pages = "062314",
    year = "2000"
}

@article{BabaeiVelni:2020wfl,
    author = "Babaei Velni, Komeil and Mohammadi Mozaffar, M. Reza and Vahidinia, M. H.",
    title = "{Evolution of entanglement wedge cross section following a global quench}",
    eprint = "2005.05673",
    archivePrefix = "arXiv",
    primaryClass = "hep-th",
    reportNumber = "IPM/P-2020/014",
    doi = "10.1007/JHEP08(2020)129",
    journal = "JHEP",
    volume = "08",
    pages = "129",
    year = "2020"
}

@article{Maldacena:1997re,
    author = "Maldacena, Juan Martin",
    title = "{The Large $N$ limit of superconformal field theories and supergravity}",
    eprint = "hep-th/9711200",
    archivePrefix = "arXiv",
    reportNumber = "HUTP-97-A097, HUTP-98-A097",
    doi = "10.4310/ATMP.1998.v2.n2.a1",
    journal = "Adv. Theor. Math. Phys.",
    volume = "2",
    pages = "231--252",
    year = "1998"
}

@article{Bao:2025psl,
    author = "Bao, Ning and Furuya, Keiichiro and Naskar, Joydeep",
    title = "{Tripartite correlation signal from multipartite entanglement of purification}",
    eprint = "2509.08209",
    archivePrefix = "arXiv",
    primaryClass = "hep-th",
    doi = "10.1007/JHEP05(2026)236",
    journal = "JHEP",
    volume = "05",
    pages = "236",
    year = "2026"
}

@article{Abajo-Arrastia:2010ajo,
    author = "Abajo-Arrastia, Javier and Aparicio, Joao and Lopez, Esperanza",
    title = "{Holographic Evolution of Entanglement Entropy}",
    eprint = "1006.4090",
    archivePrefix = "arXiv",
    primaryClass = "hep-th",
    doi = "10.1007/JHEP11(2010)149",
    journal = "JHEP",
    volume = "11",
    pages = "149",
    year = "2010"
}

@article{Balasubramanian:2011at,
    author = "Balasubramanian, V. and Bernamonti, A. and Copland, N. and Craps, B. and Galli, F.",
    title = "{Thermalization of mutual and tripartite information in strongly coupled two dimensional conformal field theories}",
    eprint = "1110.0488",
    archivePrefix = "arXiv",
    primaryClass = "hep-th",
    doi = "10.1103/PhysRevD.84.105017",
    journal = "Phys. Rev. D",
    volume = "84",
    pages = "105017",
    year = "2011"
}

@article{Balasubramanian:2010ce,
    author = "Balasubramanian, V. and Bernamonti, A. and de Boer, J. and Copland, N. and Craps, B. and Keski-Vakkuri, E. and Muller, B. and Schafer, A. and Shigemori, M. and Staessens, W.",
    title = "{Thermalization of Strongly Coupled Field Theories}",
    eprint = "1012.4753",
    archivePrefix = "arXiv",
    primaryClass = "hep-th",
    doi = "10.1103/PhysRevLett.106.191601",
    journal = "Phys. Rev. Lett.",
    volume = "106",
    pages = "191601",
    year = "2011"
}

@article{Balasubramanian:2011ur,
    author = "Balasubramanian, V. and Bernamonti, A. and de Boer, J. and Copland, N. and Craps, B. and Keski-Vakkuri, E. and Muller, B. and Schafer, A. and Shigemori, M. and Staessens, W.",
    title = "{Holographic Thermalization}",
    eprint = "1103.2683",
    archivePrefix = "arXiv",
    primaryClass = "hep-th",
    reportNumber = "HIP-2011-07-TH, UUITP-06-11",
    doi = "10.1103/PhysRevD.84.026010",
    journal = "Phys. Rev. D",
    volume = "84",
    pages = "026010",
    year = "2011"
}

@article{Allais:2011ys,
    author = "Allais, Andrea and Tonni, Erik",
    title = "{Holographic evolution of the mutual information}",
    eprint = "1110.1607",
    archivePrefix = "arXiv",
    primaryClass = "hep-th",
    doi = "10.1007/JHEP01(2012)102",
    journal = "JHEP",
    volume = "01",
    pages = "102",
    year = "2012"
}

@article{Liu:2013iza,
    author = "Liu, Hong and Suh, S. Josephine",
    title = "{Entanglement Tsunami: Universal Scaling in Holographic Thermalization}",
    eprint = "1305.7244",
    archivePrefix = "arXiv",
    primaryClass = "hep-th",
    reportNumber = "MIT-CTP/4475, MIT-CTP-4475",
    doi = "10.1103/PhysRevLett.112.011601",
    journal = "Phys. Rev. Lett.",
    volume = "112",
    pages = "011601",
    year = "2014"
}

@article{Liu:2013qca,
    author = "Liu, Hong and Suh, S. Josephine",
    title = "{Entanglement growth during thermalization in holographic systems}",
    eprint = "1311.1200",
    archivePrefix = "arXiv",
    primaryClass = "hep-th",
    reportNumber = "MIT-CTP-4510, MIT-CTP 4510",
    doi = "10.1103/PhysRevD.89.066012",
    journal = "Phys. Rev. D",
    volume = "89",
    number = "6",
    pages = "066012",
    year = "2014"
}

@article{Hartman:2013qma,
    author = "Hartman, Thomas and Maldacena, Juan",
    title = "{Time Evolution of Entanglement Entropy from Black Hole Interiors}",
    eprint = "1303.1080",
    archivePrefix = "arXiv",
    primaryClass = "hep-th",
    doi = "10.1007/JHEP05(2013)014",
    journal = "JHEP",
    volume = "05",
    pages = "014",
    year = "2013"
}

@article{Mirabi:2016elb,
    author = "Mirabi, S. and Tanhayi, M. Reza and Vazirian, R.",
    title = "{On the Monogamy of Holographic $n$-partite Information}",
    eprint = "1603.00184",
    archivePrefix = "arXiv",
    primaryClass = "hep-th",
    doi = "10.1103/PhysRevD.93.104049",
    journal = "Phys. Rev. D",
    volume = "93",
    number = "10",
    pages = "104049",
    year = "2016"
}

@article{Hayden:2011ag,
    author = "Hayden, Patrick and Headrick, Matthew and Maloney, Alexander",
    title = "{Holographic Mutual Information is Monogamous}",
    eprint = "1107.2940",
    archivePrefix = "arXiv",
    primaryClass = "hep-th",
    reportNumber = "BRX-TH-638, BRX-TH-638",
    doi = "10.1103/PhysRevD.87.046003",
    journal = "Phys. Rev. D",
    volume = "87",
    number = "4",
    pages = "046003",
    year = "2013"
}

@article{Walter:2016lgl,
    author = "Walter, Michael and Gross, David and Eisert, Jens",
    title = "{Multi-partite entanglement}",
    eprint = "1612.02437",
    archivePrefix = "arXiv",
    primaryClass = "quant-ph",
    month = "12",
    year = "2016"
}

@article{Gadde:2023zzj,
    author = "Gadde, Abhijit and Krishna, Vineeth and Sharma, Trakshu",
    title = "{Towards a classification of holographic multi-partite entanglement measures}",
    eprint = "2304.06082",
    archivePrefix = "arXiv",
    primaryClass = "hep-th",
    doi = "10.1007/JHEP08(2023)202",
    journal = "JHEP",
    volume = "08",
    pages = "202",
    year = "2023"
}

@article{araki1970entropy,
    author  = {Araki, H. and Lieb, E. H.},
    title   = {Entropy inequalities},
    journal = {Communications in Mathematical Physics},
    year    = {1970},
    volume  = {18},
    number  = {2},
    pages   = {160--170}
}

@article{Shirokov:2017zka,
    author = "Shirokov, M. E.",
    title = "{Tight continuity bounds for the quantum conditional mutual information, for the Holevo quantity and for capacities of quantum channels}",
    eprint = "1512.09047",
    archivePrefix = "arXiv",
    primaryClass = "quant-ph",
    doi = "10.1063/1.4987135",
    journal = "J. Math. Phys.",
    volume = "58",
    pages = "102202",
    year = "2017"
}

@article{Dutta:2019gen,
    author = "Dutta, Souvik and Faulkner, Thomas",
    title = "{A canonical purification for the entanglement wedge cross-section}",
    eprint = "1905.00577",
    archivePrefix = "arXiv",
    primaryClass = "hep-th",
    doi = "10.1007/JHEP03(2021)178",
    journal = "JHEP",
    volume = "03",
    pages = "178",
    year = "2021"
}

@article{Balasubramanian:2024ysu,
    author = "Balasubramanian, Vijay and Kang, Monica Jinwoo and Murdia, Chitraang and Ross, Simon F.",
    title = "{Signals of multiparty entanglement and holography}",
    eprint = "2411.03422",
    archivePrefix = "arXiv",
    primaryClass = "hep-th",
    doi = "10.1007/JHEP06(2025)068",
    journal = "JHEP",
    volume = "06",
    pages = "068",
    year = "2025"
}

@article{Balasubramanian:2025jhq,
    author = "Balasubramanian, Vijay and Jiang, Hanzhi and Ross, Simon F.",
    title = "{Time evolution of multi-party entanglement signals}",
    eprint = "2511.16729",
    archivePrefix = "arXiv",
    primaryClass = "hep-th",
    doi = "10.1007/JHEP06(2026)055",
    journal = "JHEP",
    volume = "06",
    pages = "055",
    year = "2026"
}

@article{Alishahiha:2014jxa,
    author = "Alishahiha, Mohsen and Mohammadi Mozaffar, M. Reza and Tanhayi, Mohammad Reza",
    title = "{On the Time Evolution of Holographic n-partite Information}",
    eprint = "1406.7677",
    archivePrefix = "arXiv",
    primaryClass = "hep-th",
    doi = "10.1007/JHEP09(2015)165",
    journal = "JHEP",
    volume = "09",
    pages = "165",
    year = "2015"
}

@article{Ju:2024kuc,
    author = "Ju, Xin-Xiang and Pan, Wen-Bin and Sun, Ya-Wen and Wang, Yuan-Tai and Zhao, Yang",
    title = "{More on the upper bound of holographic n-partite information}",
    eprint = "2411.19207",
    archivePrefix = "arXiv",
    primaryClass = "hep-th",
    doi = "10.1007/JHEP03(2025)184",
    journal = "JHEP",
    volume = "03",
    pages = "184",
    year = "2025"
}

@article{Ju:2024hba,
    author = "Ju, Xin-Xiang and Pan, Wen-Bin and Sun, Ya-Wen and Zhao, Yang",
    title = "{Holographic multipartite entanglement from the upper bound of $n$-partite information}",
    eprint = "2411.07790",
    archivePrefix = "arXiv",
    primaryClass = "hep-th",
    month = "11",
    year = "2024"
}

@article{Iizuka:2025ioc,
    author = "Iizuka, Norihiro and Nishida, Mitsuhiro",
    title = "{Genuine multientropy and holography}",
    eprint = "2502.07995",
    archivePrefix = "arXiv",
    primaryClass = "hep-th",
    doi = "10.1103/714c-byxq",
    journal = "Phys. Rev. D",
    volume = "112",
    number = "2",
    pages = "026011",
    year = "2025"
}

@article{Ryu:2006bv,
  author = {Ryu, Shinsei and Takayanagi, Tadashi},
  title = {Holographic Derivation of Entanglement Entropy from AdS/CFT},
  eprint = {hep-th/0603001},
  archivePrefix = {arXiv},
  primaryClass = {hep-th},
  doi = {10.1103/PhysRevLett.96.181602},
  journal = {Phys. Rev. Lett.},
  volume = {96},
  pages = {181602},
  year = {2006}
}

@article{Ryu:2006ef,
  author = {Ryu, Shinsei and Takayanagi, Tadashi},
  title = {Aspects of Holographic Entanglement Entropy},
  eprint = {hep-th/0605073},
  archivePrefix = {arXiv},
  primaryClass = {hep-th},
  doi = {10.1088/1126-6708/2006/08/045},
  journal = {JHEP},
  volume = {08},
  pages = {045},
  year = {2006}
}

@article{Hubeny:2007xt,
  author = {Hubeny, Veronika E. and Rangamani, Mukund and Takayanagi, Tadashi},
  title = {A Covariant Holographic Entanglement Entropy Proposal},
  eprint = {0705.0016},
  archivePrefix = {arXiv},
  primaryClass = {hep-th},
  doi = {10.1088/1126-6708/2007/07/062},
  journal = {JHEP},
  volume = {07},
  pages = {062},
  year = {2007}
}

@article{Zou:2020bly,
    author = "Zou, Yijian and Siva, Karthik and Soejima, Tomohiro and Mong, Roger S. K. and Zaletel, Michael P.",
    title = "{Universal tripartite entanglement in one-dimensional many-body systems}",
    eprint = "2011.11864",
    archivePrefix = "arXiv",
    primaryClass = "quant-ph",
    doi = "10.1103/PhysRevLett.126.120501",
    journal = "Phys. Rev. Lett.",
    volume = "126",
    number = "12",
    pages = "120501",
    year = "2021"
}

@article{Harper:2024ker,
    author = "Harper, Jonathan and Takayanagi, Tadashi and Tsuda, Takashi",
    title = "{Multi-entropy at low Renyi index in 2d CFTs}",
    eprint = "2401.04236",
    archivePrefix = "arXiv",
    primaryClass = "hep-th",
    reportNumber = "YITP-24-02",
    doi = "10.21468/SciPostPhys.16.5.125",
    journal = "SciPost Phys.",
    volume = "16",
    number = "5",
    pages = "125",
    year = "2024"
}

@article{Iizuka:2025caq,
    author = "Iizuka, Norihiro and Lin, Simon and Nishida, Mitsuhiro",
    title = "{More on genuine multientropy and holography}",
    eprint = "2504.16589",
    archivePrefix = "arXiv",
    primaryClass = "hep-th",
    doi = "10.1103/x76v-mr6n",
    journal = "Phys. Rev. D",
    volume = "112",
    number = "6",
    pages = "066014",
    year = "2025"
}

@article{Gadde:2022cqi,
    author = "Gadde, Abhijit and Krishna, Vineeth and Sharma, Trakshu",
    title = "{New multipartite entanglement measure and its holographic dual}",
    eprint = "2206.09723",
    archivePrefix = "arXiv",
    primaryClass = "hep-th",
    reportNumber = "TIFR/TH/22-34",
    doi = "10.1103/PhysRevD.106.126001",
    journal = "Phys. Rev. D",
    volume = "106",
    number = "12",
    pages = "126001",
    year = "2022"
}

@article{Akers:2019gcv,
    author = "Akers, Chris and Rath, Pratik",
    title = "{Entanglement Wedge Cross Sections Require Tripartite Entanglement}",
    eprint = "1911.07852",
    archivePrefix = "arXiv",
    primaryClass = "hep-th",
    doi = "10.1007/JHEP04(2020)208",
    journal = "JHEP",
    volume = "04",
    pages = "208",
    year = "2020"
}

@article{Hayden:2021gno,
    author = "Hayden, Patrick and Parrikar, Onkar and Sorce, Jonathan",
    title = "{The Markov gap for geometric reflected entropy}",
    eprint = "2107.00009",
    archivePrefix = "arXiv",
    primaryClass = "hep-th",
    doi = "10.1007/JHEP10(2021)047",
    journal = "JHEP",
    volume = "10",
    pages = "047",
    year = "2021"
}

@article{Takayanagi:2017knl,
    author = "Takayanagi, Tadashi and Umemoto, Koji",
    title = "{Entanglement of purification through holographic duality}",
    eprint = "1708.09393",
    archivePrefix = "arXiv",
    primaryClass = "hep-th",
    reportNumber = "YITP-17-89, IPMU17-0115",
    doi = "10.1038/s41567-018-0075-2",
    journal = "Nature Phys.",
    volume = "14",
    number = "6",
    pages = "573--577",
    year = "2018"
}

@article{Nguyen:2017yqw,
    author = "Nguyen, Phuc and Devakul, Trithep and Halbasch, Matthew G. and Zaletel, Michael P. and Swingle, Brian",
    title = "{Entanglement of purification: from spin chains to holography}",
    eprint = "1709.07424",
    archivePrefix = "arXiv",
    primaryClass = "hep-th",
    doi = "10.1007/JHEP01(2018)098",
    journal = "JHEP",
    volume = "01",
    pages = "098",
    year = "2018"
}

@article{Basak:2023uix,
    author = "Basak, Jaydeep Kumar and Giataganas, Dimitrios and Mondal, Sayid and Wen, Wen-Yu",
    title = "{Reflected entropy and Markov gap in noninertial frames}",
    eprint = "2306.17490",
    archivePrefix = "arXiv",
    primaryClass = "quant-ph",
    doi = "10.1103/PhysRevD.108.125009",
    journal = "Phys. Rev. D",
    volume = "108",
    number = "12",
    pages = "125009",
    year = "2023"
}

@article{Chen:2026xtx,
    author = "Chen, Xiantong and Ji, Xuanting and Sun, Ya-Wen",
    title = "{Multipartite entanglement characterizing topological phase transitions in holographic nodal line semimetals}",
    eprint = "2602.01545",
    archivePrefix = "arXiv",
    primaryClass = "hep-th",
    doi = "10.1007/JHEP07(2026)072",
    journal = "JHEP",
    volume = "07",
    pages = "072",
    year = "2026"
}

@article{Terhal:2002riz,
    author = "Terhal, Barbara M. and Horodecki, Michal and Leung, Debbie W. and DiVincenzo, David P.",
    title = "{The entanglement of purification}",
    eprint = "quant-ph/0202044",
    archivePrefix = "arXiv",
    doi = "10.1063/1.1498001",
    journal = "J. Math. Phys.",
    volume = "43",
    number = "9",
    pages = "4286--4298",
    year = "2002"
}

@article{Ju:2023tvo,
    author = "Ju, Xin-Xiang and Lai, Teng-Zhou and Sun, Ya-Wen and Wang, Yuan-Tai",
    title = "{Holographic n-partite information in hyperscaling violating geometry}",
    eprint = "2304.11430",
    archivePrefix = "arXiv",
    primaryClass = "hep-th",
    doi = "10.1007/JHEP08(2023)064",
    journal = "JHEP",
    volume = "08",
    pages = "064",
    year = "2023"
}

@article{Ju:2025tgg,
    author = "Ju, Xin-Xiang and Sun, Ya-Wen and Zhao, Yang",
    title = "{Upper bound of holographic entanglement entropy combinations}",
    eprint = "2505.11059",
    archivePrefix = "arXiv",
    primaryClass = "hep-th",
    doi = "10.1007/JHEP09(2025)085",
    journal = "JHEP",
    volume = "09",
    pages = "085",
    year = "2025"
}

@article{Fujiki:2026qdt,
    author = "Fujiki, Kosei and Tasuki, Kenya",
    title = "{Multi-entropy in heavy local quenches}",
    eprint = "2606.12526",
    archivePrefix = "arXiv",
    primaryClass = "hep-th",
    reportNumber = "YITP-26-65",
    month = "6",
    year = "2026"
}

@article{Popescu:2006rhr,
    author = "Popescu, Sandu and Short, Anthony J. and Winter, Andreas",
    title = "{Entanglement and the foundations of statistical mechanics}",
    eprint = "quant-ph/0511225",
    archivePrefix = "arXiv",
    doi = "10.1038/nphys444",
    journal = "Nature Phys.",
    volume = "2",
    number = "11",
    pages = "754--758",
    year = "2006"
}

@article{Goldstein:2005aib,
    author = "Goldstein, Sheldon and Lebowitz, Joel L. and Tumulka, Roderich and Zanghi, Nino",
    title = "{Canonical Typicality}",
    eprint = "cond-mat/0511091",
    archivePrefix = "arXiv",
    doi = "10.1103/PhysRevLett.96.050403",
    journal = "Phys. Rev. Lett.",
    volume = "96",
    pages = "050403",
    year = "2006"
}

@article{Greenberger:1990uox,
    author = "Greenberger, Daniel M. and Horne, Michael A. and Shimony, Abner and Zeilinger, Anton",
    title = "{Bell{\textquoteright}s theorem without inequalities}",
    doi = "10.1119/1.16243",
    journal = "Am. J. Phys.",
    volume = "58",
    number = "12",
    pages = "1131",
    year = "1990"
}

\end{document}